\documentclass[12pt]{article}

\usepackage{times}
  
\newenvironment{sciabstract}{%
\begin{quote} \bf}
{\end{quote}}

\newcounter{lastnote}

\title{Rapid formation of supermassive black hole binaries in galaxy mergers with gas}

\author
{L.~Mayer,$^{1,2\ast}$ S.~Kazantzidis,$^{3\ast}$ P. Madau,$^{4,5}$\\
M.Colpi,$^6$ T.Quinn,$^7$ J.~Wadsley$^8$\\
\\
\normalsize{$^1$Institut f\"ur Astronomie, ETH Z\"urich, Wolfgang-Pauli-Strasse 16,}\\
\normalsize{CH-8093 Z\"urich, Switzerland.} \\
\normalsize{$^2$Institute for Theoretical Physics, University of Zurich,}\\
\normalsize{Winterthurestrasse 190, CH-8057 Z\"urich, Switzerland.}\\
\normalsize{$^3$Kavli Institute for Particle Astrophysics and Cosmology, Department of Physics,}\\
\normalsize{Stanford University, P.O. Box 20450, MS 29, Stanford,CA 94309 USA.}\\
\normalsize{$^4$Department of Astronomy, University of California at Santa Cruz,}\\
\normalsize{1156 High Street, Santa Cruz, CA 95064, USA.}\\
\normalsize{$^5$Max Planck Institute f\"ur Astrophysik, Karl-Schwarzschild Strasse 1,}\\
\normalsize{85740, Garching bei Muenchen, Germany.}\\
\normalsize{$^6$Dipartimento di Fisica, Universit\`a di Milano Bicocca,}\\
\normalsize{Piazza della Scienza 3. I-20126 Milano, Italy.}\\
\normalsize{$^7$Department of Astronomy, University of Washington,}\\
\normalsize{Stevens Way, Seattle, WA 98195, USA.}\\
\normalsize{$^8$Department of Physics and Astronomy, McMaster University,}\\
\normalsize{Hamilton, ON L8S 4M1, Canada.}\\
\\
\normalsize{$^\ast$To whom correspondence should be addressed; E-mail:} \\
\normalsize{lucio@phys.ethz.ch; stelios@slac.stanford.edu.}
}

\date{}

\begin{document} 

\baselineskip24pt

\maketitle 

\begin{sciabstract}

Supermassive black holes (SMBHs) are a ubiquitous component of the nuclei 
of galaxies. It is normally assumed that, following the merger of two massive galaxies, a
SMBH binary will form, shrink due to stellar or gas dynamical processes
and  ultimately coalesce by emitting a burst of gravitational waves.
However, so far it has not been possible to show how two SMBHs bind
during a galaxy merger with gas due to the difficulty of modeling a wide range 
of spatial scales. Here we report hydrodynamical 
simulations that track the formation of a SMBH binary down to scales of a 
few light years following the collision between two spiral galaxies. A
massive, turbulent nuclear gaseous disk arises as a
result of the galaxy merger. The black holes form an eccentric 
binary in the disk in less than a million years as a result of the
gravitational drag from the gas rather than from the stars.

\end{sciabstract}

\maketitle

Supermassive black holes (SMBHs) weighting up to a billion solar masses are thought
to reside at the center of all massive galaxies {\it(1-3)}.
According to the standard paradigm of structure formation in the Universe, galaxies
merge frequently as their dark matter halos assemble in a hierarchical fashion {\it(4,5)}.
As SMBHs become incorporated into progressively larger halos, 
they sink to the center of the more massive progenitor owing to dynamical friction and eventually 
form a binary {\it(5-8)}. In a purely stellar background, as the binary separation 
decays, the effectiveness of dynamical friction slowly declines, and the pair then becomes
tightly bound via 
three-body interactions, namely by capturing stars that pass close to the 
holes and ejecting them at much higher velocities {\it(5-7)}. 
If the hardening continues sufficiently far, the loss of orbital energy due to gravitational 
wave emission finally takes over, and the two SMBHs coalesce in less than a Hubble time. 
But the binary may stop sinking before gravitational 
radiation becomes important since there is a finite supply of stars on intersecting orbits
{\it(5,9)}.

During the assembly of galaxies, however, their SMBHs likely evolve 
within gas-rich systems. Merging systems like the Ultraluminous 
Infrared Galaxies (ULIRGs) NGC6240 and Arp220 harbor large 
concentrations of gas, in excess of $10^9$ M$_{\odot}$, at their 
center, in the form of either a turbulent irregular structure or of 
a kinematically coherent, rotating disk{\it(10-12)}. Massive 
rotating nuclear disks of molecular gas are also ubiquitous in galaxies
that appear to have just undergone a major merger, such as Markarian 231{\it(13)}. 

Gas dynamics may profoundly affect the pairing of SMBHs both during and 
after their host galaxies merge{\it(14-18)}. Recent simulations of the orbital evolution of 
SMBHs within an equilibrium, rotationally-supported, gaseous disk have 
shown that friction against the gaseous background leads to the 
formation of a tightly bound SMBH binary with final separation 
$<1$ pc in about $10^7$ yr {\it(16,17)}. 
Yet such simulations begin with 
ad hoc initial conditions, with the black holes already forming a loosely bound pair, 
while in reality the orbital configuration of the black holes and the structure
and thermodynamics of the nuclear region, which can affect the 
drag {\it(17-18)}, will be the end result of the complex gravitational and 
hydrodynamical processes involved in the merger. How a pair of SMBHs 
binds in a dissipational galaxy merger is thus still unclear.

Here we report on high resolution N-body + smoothed particle
hydrodynamics (SPH) simulations of mergers between galaxies with SMBHs having enough 
dynamic range to follow the holes from a hundred kiloparsecs down to parsec 
scales, bridging about ten orders of magnitude in density. 
We start with two equal-mass galaxies similar to the Milky Way, comprising a disk of stars and gas
with a surface density distribution that follows an exponential law, 
a stellar bulge, and a massive and extended spherical dark matter halo 
whose mass, radius, density profile and angular momentum is consistent with current structure formation 
models {\it(19)}. Their initial orbit
is parabolic and their distance of closest approach is 50 kpc, consistent with typical
values found in cosmological simulations of structure formation{\it(20)}. 
A particle of mass $2.6 \times 10^6 M_{\odot}$ is placed at the
center of each bulge to represent a SMBH. The simulations include
radiative cooling and star formation {\it(14)} and have a spatial resolution 
of $100$ pc {\it(19)}. The computational volume is refined during the late stage
of the merger with the technique of particle splitting {\it(19)}, achieving a spatial resolution 
of $2$ pc with as many as $2\times 10^6$ gas particles within the nuclear region.

Initially the separation of the two black holes evolves as that of the two 
gaseous cores in which they are embedded. The galaxies approach each other
several times as they sink into one another via dynamical friction. After about 5 Gyr the 
dark matter halos have nearly merged and the two baryonic cores, separated by 
about 6 kpc, continue to spiral down (Figure 1).
As much as 60\% of the gas originally present in the 
galaxies has been funneled to the inner few hundred parsecs of each core by 
tidal torques and shocks occurring in the repeated fly-bys between the two 
galaxies {\it(14, 21,22)} (Figure 1).
Each of the two SMBHs is embedded in a  
rotating gaseous disk of mass $\sim 4 \times 10^8$ M$_{\odot}$ and size 
of a few hundred parsecs, produced by such gas inflow. 
At this stage we stop the simulation and we restart it with increased resolution {\it(19)}. 

\begin{figure}
\vskip 8.0cm 
{\includegraphics{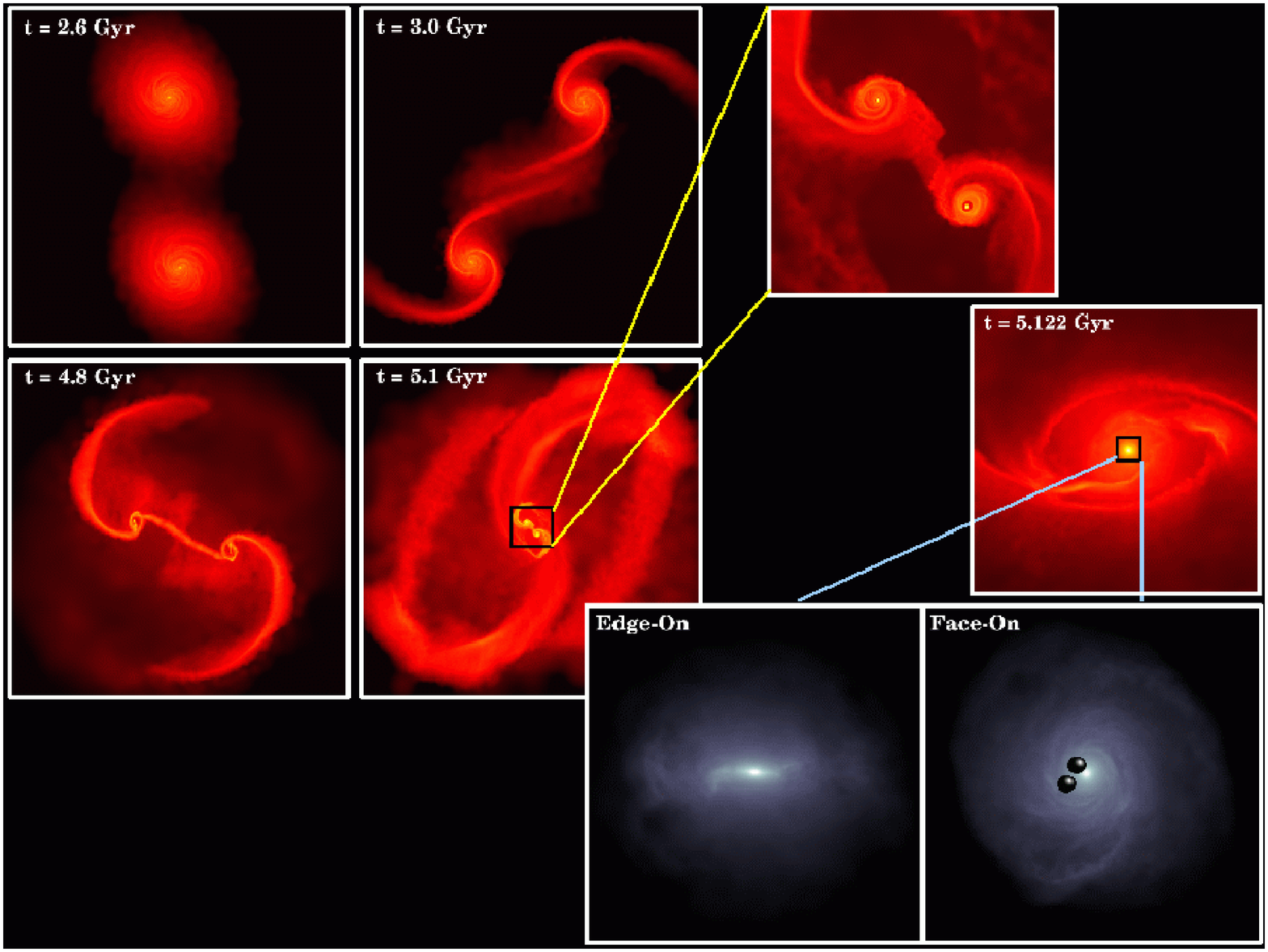}}
\caption[]{\small The different stages of the merger between two identical
disk galaxies. The color coded density maps of the gas component  
are shown using a logarithmic scale, with brighter colors for higher densities. 
The four panels to the left show the large-scale 
evolution at different times. The boxes are 120 kpc on a side (top) 
and 60 kpc on a side (bottom) and the density ranges between $10^{-2}$ atoms
cm$^{-3}$ and  $10^{2}$ atoms cm$^{-3}$. During the interaction tidal forces tear the 
galactic disks apart, generating spectacular tidal tails and plumes.
The panels to the right show a zoom in 
of the very last stage of the merger, about 100 million years before
the two cores have fully coalesced   (upper panel), and 2 million years after 
the merger (middle panel), when a massive, rotating nuclear gaseous
disk embedded in a series of large-scale ring-like structures has formed.  
The boxes are now 8 kpc on a side and the density ranges between $10^{-2}$ atoms
cm$^{-3}$ and  $10^{5}$ atoms cm$^{  -3}$.  
The two bottom panels, with a grey color scale, show the detail of the 
inner 160 parsecs of the middle panel; the nuclear disk is shown 
edge-on (left) and face-on (right), and the two black holes are also shown 
in the face-on image. An equation of state with $\gamma=7/5$ was used in the refined part 
of the simulation.}
\end{figure}

The radiation physics in the refined simulation is modeled via an 
effective equation of state that accounts for the net balance of 
radiative heating and cooling. In a previously performed non-refined simulation
a starburst with a peak star formation rate of $\sim 30$ M$_{\odot}$/yr takes 
place when the cores finally merge {\it(14)}.
We do not account for any conversion of gas into stars in the refined 
simulation to limit the computational burden: this will not affect our 
conclusions since we explore a phase lasting $< 10^7$ yr after the merger, 
namely much shorter than the duration of the starburst in the 
non-refined simulation, which is close to $10^8$ yr {\it(19)}.
Calculations that include radiative transfer show that the 
thermodynamic state of a solar metallicity gas heated by a starburst 
can be well approximated by an ideal gas with 
adiabatic index $\gamma=1.3-7/5$ over a wide range of densities {\it(23,24)}. 
We assume $\gamma=7/5$ and 
include the irreversible heating generated by shocks via an artificial 
viscosity term in the internal energy equation {\it(19)}.

The gaseous cores finally merge at $t \sim 5.12$ Gyr, forming a single nuclear 
disk  with a mass of $3\times 10^9 M_{\odot}$ and a size of $\sim 75$ pc. The two
SMBHs are now embedded in such nuclear disk. The disk is more massive than 
the sum of the two progenitor nuclear disks formed earlier because 
further gas inflow occurs in the last stage of the galaxy collision.
It is surrounded by several rings and by a more 
diffuse, rotationally-supported envelope extending out to more than
a kiloparsec from the center (Figure 1). A background of dark matter and stars 
distributed in a spheroid is also present but the gas component is
dominant in mass within a few hundred pc from the center. From now on the orbital 
decay of the holes is dominated by
dynamical friction against the gaseous disk. The black holes are 
on eccentric orbits (the eccentricity is $e \sim 0.5$
where $e=(r_{\rm apo} - r_{\rm peri})/(r_{\rm apo} + r_{\rm peri})$, $r_{\rm apo}$ and $r_{\rm peri}$ being,
respectively, the apocenter and pericenter of the orbit) near the plane of the disk
{\it(19)}, and move at a speed 
$v_{\rm BH} \sim 200-300$ km s$^{-1}$ relative to the disk's center of mass.
The typical ambient sound speed is $v_s \sim 45$  km s$^{-1}$,
a legacy of the strong shock heating occurring as the galaxy cores 
merge. The disk is rotationally supported, $v_{\rm rot} \sim 300$ km s$^{-1}$, but 
is also highly turbulent, having a typical velocity dispersion $v_{\rm turb} \sim 100$ km s$^{-1}${\it(19)}. 
Its scale height, $\sim 20$ pc, and typical density, $10^3-10^4$ atoms cm$^{-3}$, 
are comparable to those of observed nuclear disks {\it(11)}.

\begin{figure}
\vskip 7.5cm
{\includegraphics{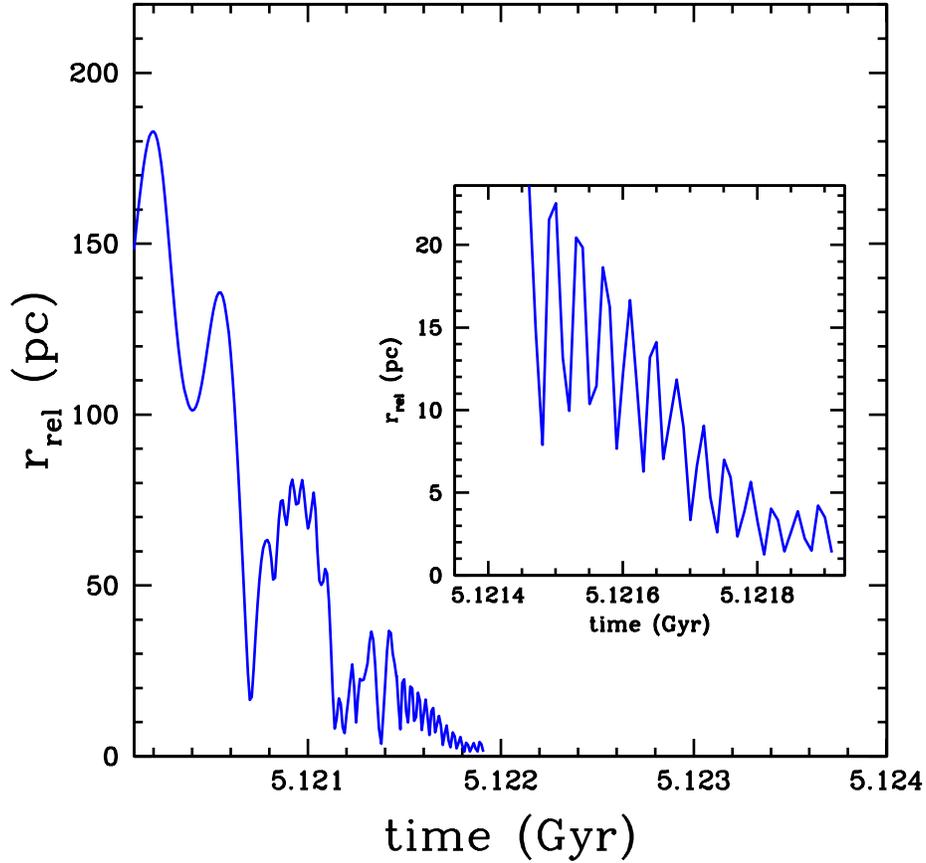}}
\caption[]{\small Orbital separation of the two black holes as a function 
of time during the last stage of the galaxy merger shown in Figure 1.
The orbit of the pair is eccentric until the end of the simulation. The two peaks at scales of 
tens of parsecs at around $t=5.1213$ Gyr mark the end of the phase during which
the two holes are still embedded in two distinct gaseous cores. Until this 
point the orbit is the result of the relative motion of the cores combined
with the relative motion of each black hole relative to the surrounding core, explaining 
the presence of more than one orbital frequency. The inset shows the details 
of the last part of the orbital evolution, which takes place in the nuclear 
disk arising from the merger of the two cores. The binary stops 
shrinking when the separation approaches the softening length 
(2 pc).}
\end{figure}

The two SMBHs sink down from about 40 pc to a few parsecs, our resolution limit, in less than 
a million years (Figure 2). At 
this point the two  holes are gravitationally bound to each other, as 
the mass of the gas enclosed within their separation is less than 
the mass of the binary. The gas controls the orbital decay, not the stars.
Dynamical friction against the stellar background would bring the two
black holes this close only on a much longer timescale, $\sim 5 \times 10^7$ yr 
{\it(19)}.
A short sinking timescale due to the gas is expected because of the high gas densities and since 
the decay occurs in the supersonic regime {\it(19)}, being $v_{\rm BH} > v_{\rm turb} > v_{s}$. 
The subsequent hardening of the binary will depend on the details of gasdynamics and other 
processes at scales below our resolution {\it(16-19)}.

If radiative cooling is completely suppressed during the merger, for example as a result of radiative heating 
following gas accretion onto the SMBHs, the gas would evolve adiabatically ($\gamma=5/3$). In this case 
the hardening process is significantly slowed 
down, and gas and stars contribute similarly to the drag {{\it(19)}}.
However, if the SMBHs become active only after the nuclear disk arises
and keep accreting the surrounding gas at the Eddington limit until the binary forms, their radiative heating should not be enough to alter significantly the
energy balance implicitly assumed in the $\gamma=7/5$ simulation {${\it (19)}$}. 

Here we have considered a merger between galaxies in which the gas accounts for only $10\%$ of the disk 
mass, a typical gas fraction in present-day spirals. Much larger gas fractions should be 
common at high redshift, when most of the merger activity takes place and 
massive galaxies have just begun to assemble their 
stellar component{\it(25)}. Even more massive and denser nuclear disks should 
form then and, since dynamical friction is proportional to the density of 
the background {\it(16,19)}, a pair of SMBHs will bind even faster than
in our calculations.
Coalescing SMBH binaries will thus
be common at high redshift and are among the primary candidate sources of
gravitational waves at mHz frequencies, the range probed by the 
space-based Laser Interferometer Space Antenna (LISA){\it(7,26)}. 
Moreover, even at the present epoch a typical bright galaxy
has a more massive stellar bulge relative to our models, and 
hence harbors more massive SMBHs {\it(1-3)} that will decay faster because
dynamical friction is stronger for larger bodies.

Three-body encounters between ambient stars and
a SMBH binary may deplete the nuclear region and turn a stellar cusp into 
a low-density core at scales of tens of parsecs {\it(27)}. This would explain why the brightest ellipticals, 
very likely the end result of several mergers, have shallow stellar cores 
{\it (28, 29)}. In our scenario the orbital decay is driven by the gas rather than by 
the stellar background and occurs 
on such a short timescale that the interaction between the binary and the stellar spheroid would be 
negligible and should hardly affect the stellar density profile. Gas-rich mergers yield steep 
stellar profiles owing to the dramatic gas inflow and subsequent star formation {\it(21)}.
We expect that such
cuspy profiles will be preserved in the remnant because of the negligible interaction between
the binary SMBHs and the stars implied by our calculations.
Remnants of dissipational mergers such as Markarian 231 {\it(13)} do 
indeed exhibit a steep stellar profile at least down to a hundred parsecs, and a small effective radius 
reminiscent of that of low-luminosity ellipticals galaxies that have cuspy profiles down
to a few parsecs {\it(28)}.
Our prediction can be thoroughly tested with current and future high resolution 
multi-wavelength observations capable of probing the inner few parsecs of the remnants of dissipational mergers.

\bigskip

{\large{\bf References and Notes}}

\begin{enumerate}

\item Kormendy, J. \& Richtsone, D., {\it Ann. Rev. Astron. Astrophys.} \textbf{33}, 581 (1995)

\item Richtsone, D., {\it et al.}, {\it Nature}, {\textbf 395}, A14 (1998)

\item  Tremaine, S. et al., {\it Astrophys. J.}, \textbf{574}, 740 (2002)

\item  Springel, V., {\it et al.}, {\it Nature}, \textbf{435}, 629 (2006)

\item  Volonteri, M., Haardt, F. \& Madau, P., {\it Astrophys. J.}, \textbf{582}, 559 (2003)

\item  Begelman, M. C., Blandford, R. D. \& Rees, M. J., {\it Nature} \textbf{287}, 307, (1980)

\item  Milosavljevic, M. \& Merritt, D.,{\it Astrophys. J.}, \textbf{563}, 34 (2001)

\item Sesana, A., Haardt, F., Madau., P. \& Volonteri, M., {\it Astrophys. J.}, \textbf{611}, 23 (2005)

\item  Berczik, P., Merritt, D., Spurzem, R., \& Bischof, H., {\it Astrophys. J}. \textbf{633}, 680, (2005)

\item  Greve, T.R., Papadopoulos, P.P., Gao, Y., \& Radford, S.J.E., submitted to Astrophys. J. (2006) (astro-ph/0610378)

\item  Downes, D. \& Solomon, P. M.,{\it Astrophys. J.}, \textbf{507}, 615 (1998)

\item  Davies, R. I.,  Tacconi, L. J.  \& Genzel, R.{\it Astrophys. J.}, \textbf{602}, 148 (2004)

\item  Davies, R. I.,  Tacconi, L. J.  \& Genzel, R. {\it Astrophys. J.}, \textbf{613}, 781 (2004)

\item  Kazantzidis, S., Mayer, L., Colpi, M., Madau, P., Debattista, V., Quinn, T., Wadsley, J. \& Moore, B.,{\it Astrophys. J.}, \textbf{623}, L67 (2005)

\item  Escala, A., Larson, R. B., Coppi, P. S., \& Mardones, D., {\it Astrophys. J.}, \textbf{607}, 765 (2004)

\item  Escala A., Larson, R. B., Coppi, P. S. \& Mardones, D., {\it Astrophys. J.}, \textbf{630}, 152 (2005)

\item  Dotti, M., Colpi, M. \& Haardt, F.{\it Mon. Not. R. Astron. Soc.}, \textbf{367}, 103 (2006)

\item  Ostriker, E., {\it Astrophys. J.}, \textbf{513}, 252 (1999)

\item  Material and Methods are available as supporting material on {\it Science} online.

\item Khochfar, S. \& Burkert, A., {\it Astron. Astrophys.}, \textbf{445}, 403 (2006)

\item  Barnes, J. \& Hernquist, L., {\it Astrophys. J.}, \textbf{471}, 115 (1996)

\item  Springel, V., Di Matteo, T., \& Hernquist, L., {\it Mon. Not. R. Astron. Soc.}, \textbf{361}, 776 (2005)

\item  Spaans, M. \& Silk, J., {\it Astrophys. J.}, \textbf{538}, 115 (2000)

\item  Klessen, R.S., Spaans, M., Jappsen, A., {\it Mon. Not. R. Astron. Soc.}, \textbf{374}, L29 (2007)

\item  Genzel, R., {\it et al.}, {\it Nature}, \textbf{442}, 786 (2005)

\item  Cutler, C. \& Thorne, K.S., {\it Proceedings of GR16} (Durban, South Africa) (2002) 

\item  Merritt, D., {\it Astrophys. J.}, {\textbf 648}, 976 (2006)

\item  Lauer, T.R. {\it et al.},{\it Astronom. J.},  {\textbf 110}, 2622 (1995)

\item  Graham, A. W., {\it Astrophys. J.}, {\textbf 613}, L33 (2004)

\item  We acknowledge discussions with  Marcella Carollo, Massimo Dotti, Andres Escala, 
Savvas Koushiappas, David Merritt, Rainer Spurzem, Monica Valluri, and Marta Volonteri
S.Kazantzidis is funded by the U.S. Department of Energy through a KIPAC Fellow
ship at Stanford  University and the Stanford Linear Accelerator Center.
P. Madau acknowledges support by NASA and by the Alexander von Humboldt Foundation.
All simulations were performed on Lemieux at the Pittsburgh Supercomputing 
Center, on the Zbox and Zbox2 supercomputers at the University of Z\"urich,
and on the Gonzales cluster at ETH Z\"urich.

\end{enumerate}

\smallskip

{\bf Supporting Online Material} 

Material and Methods

SOM text

Figs:S1 to S5

\newpage

%%%%%%%%%%%%%%%%%%%%%%%%%%%%%%%
%% SUPPLEMENTARY INFORMATION %%
%%%%%%%%%%%%%%%%%%%%%%%%%%%%%%%

\begin{center}
\vspace*{-10.00pt}
{\Large \bf SUPPORTING ONLINE MATERIAL} 
\end{center}

\noindent Here we briefly describe the setup of the initial conditions and the 
numerical methods used to perform the simulations presented
in the Letter along with resolution tests.
This is followed by a critical discussion of the assumptions behind
the modeling of thermodynamics in the simulations. With the aid of
additional numerical experiments we also explore how the structure
of the nuclear region and the sinking time of the black holes depend
on thermodynamics.

\section{Numerical Methods}

\subsection{The N-Body+SPH code:GASOLINE}

We have used the fully parallel, N-Body+smoothed particle hydrodynamics (SPH) 
code GASOLINE to compute the evolution of both the collisionless and 
dissipative component in the simulations. A detailed description of the code is available in the literature${\it(S1)}$. 
Here we recall its essential features. 
GASOLINE computes gravitational forces 
using a tree--code${\it(S2)}$ that employs multipole expansions to approximate the gravitational
acceleration on each particle. A tree is built with each node storing
its multipole moments.  Each node is recursively divided into smaller
subvolumes until the final leaf nodes are reached.  Starting from the
root node and moving level by level toward the leaves of the tree, we
obtain a progressively more detailed representation of the underlying
mass distribution. In calculating the force on a particle, we can
tolerate a cruder representation of the more distant particles leading
to an $O(N \log{N})$ method. Since we only need a crude representation 
for distant mass, the concept of ``computational locality'' translates 
directly to spatial locality and leads to a natural domain decomposition.  
Time integration is carried out using the leapfrog method, which is a 
second-order symplectic integrator requiring only one costly force 
evaluation per timestep and only one copy of the physical state of the system.

\begin{figure}
\vskip 9.0cm 
{\includegraphics{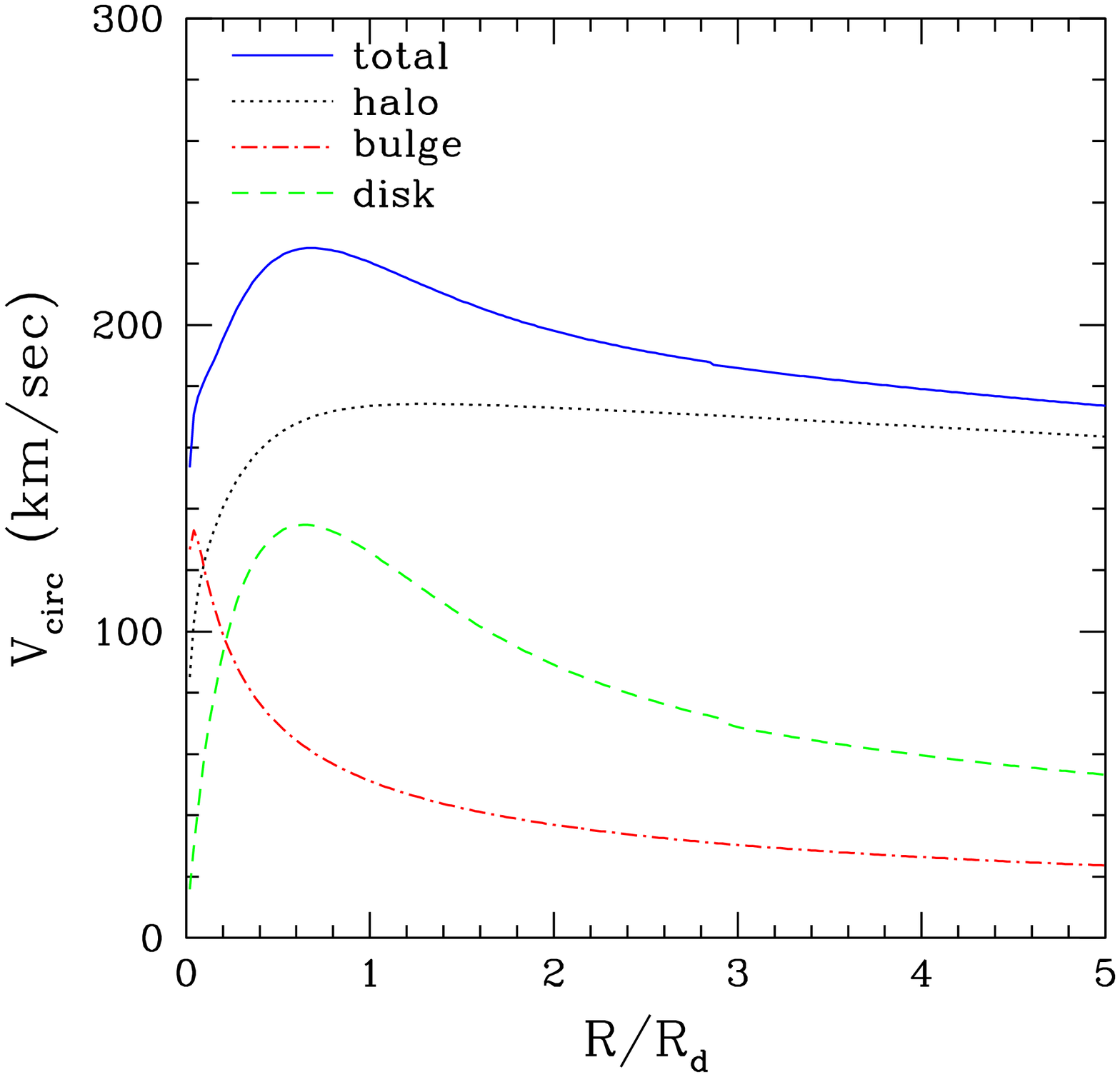}}
\smallskip {\small Figure S1. Rotation curve of the multi-component galaxy model used in the merger
simulation. The different lines represent the contribution of the different component of the galaxy to the total rotation curve (blue line) as indicated in the Figure.}
\end{figure}

SPH is a technique of using particles to integrate 
fluid elements representing gas${\it(S3,S4)}$
GASOLINE is fully Lagrangian, spatially and temporally adaptive and efficient
for large $N$. It employs
radiative cooling in the galaxy merger simulation used as a starting point for the refined
simulations presented in this Report. We use a standard cooling 
function for a primordial mixture of atomic hydrogen and helium. We shut off radiative 
cooling at temperatures below $2 \times 10^{4}$ K that is 
about a factor of $2$ higher than the temperature at which atomic radiative 
cooling would drop sharply due to the adopted cooling function. 
With this choice we take into account non-thermal, turbulent pressure to model 
the warm ISM of a real galaxy${\it(S5)}$. Unless strong shocks occur (this will be
the case during the final stage of the merger) the gaseous disk evolves nearly isothermally
since radiative cooling is very efficient at these densities ($< 100$ atoms/cm$^3$) and 
temperatures ($10^4$ K), and thus dissipates rapidly the compressional heating resulting
from the non-axisymmetric structures (spiral arms, bars) that soon develop in each
galaxy as a result of self-gravity and the tidal disturbance of the companion. 
The cooling rate would increase with the inclusion of metal lines, but ${\it(S31)}$
have shown that the equation of state of gas at these densities is still nearly isothermal
($\gamma \sim 0.9-1.1$) for a range of metallicities (with $\gamma$ being lower for higher
metallicity), supporting the validity of simple choice for the cooling function.
Cooling by metals will surely be important below $10^4$ K, but this would be irrelevant
in our scheme since we have imposed a temperature floor of $2 \times 10^4$ K to account 
for non-thermal pressure (see above). The specific internal 
energy of the gas is integrated using the asymmetric formulation. With this formulation 
the total energy is conserved exactly (unless physical dissipation due to cooling
processes is included) and entropy is closely conserved away from shocks,
which makes it similar to alternative entropy integration approaches${\it(S6)}$.
Dissipation in shocks is modeled using the 
quadratic term of the standard Monaghan artificial viscosity${\it(S4)}$. 
The Balsara correction term is used to reduce unwanted shear viscosity${\it(S7)}$.
The galaxy merger simulation${\it(S8)}$ includes star formation as well.
The star formation algorithm is such that gas particles in dense, cold Jeans unstable regions and in 
convergent flows spawn star particles at a rate proportional to the local dynamical
time${\it(S9,S10)}$. The star formation efficiency was set to $0.1$,
which yields a star formation rate of $1-2 M_{\odot}$/yr for models in isolation that have a 
disk gas mass and surface density comparable to those of the Milky Way. 

\subsection{The simulations of galaxy mergers}

For the Report we performed a refined calculation of a galaxy merger simulation
between two identical galaxies.
The initial conditions of this and other similar merger simulations are described in a
previous paper${\it(S8)}$.
We employed a multicomponent galaxy model constructed using the technique originally developed 
in ${\it(S11,S12)}$, its structural parameters being consistent with the $\Lambda$CDM paradigm for structure formation${\it(S13)}$.
The model comprises a spherical and isotropic Navarro-Frenk-and-White (NFW)
dark matter (DM) halo${\it(S14,S15)}$, an exponential disk, and 
a spherical,  non-rotating bulge. We adopted parameters 
from the Milky Way model A1 of ${\it(S16)}$. Specifically, the DM
halo has a virial mass of $M_{\rm vir}=10^{12} M_{\odot}$, a
concentration parameter of $c=12$, and a dimensionless spin parameter
of $\lambda=0.031$. The mass, thickness and resulting scale length of the disk
are $M_{\rm d}=0.04 M_{\rm vir}$, $z_{0}=0.1 R_{\rm d}$, and $R_{\rm d}=3.5$ kpc, 
respectively. The bulge mass and scale radius are $M_{\rm b}=0.008 M_{\rm vir}$
and $a=0.2 R_{\rm d}$, respectively. The DM halo was adiabatically contracted 
to respond to the growth of the disk and bulge${\it(S17)}$ resulting
in a model with a central total density slope close to isothermal. 
The galaxies are consistent with the stellar mass Tully-Fisher and size-mass relations.
To each of them we add a (softened) particle initially at rest at the center of the bulge 
to represent a supermassive black hole (SMBH). We used a SMBH mass equal to 
$M_{\rm BH}=2.6 \times 10^{6} M_{\odot}$, consistent with the $M_{BH}-\sigma$ relation${\it(S8)}$.

\begin{figure}
\vskip 9cm 
{\includegraphics{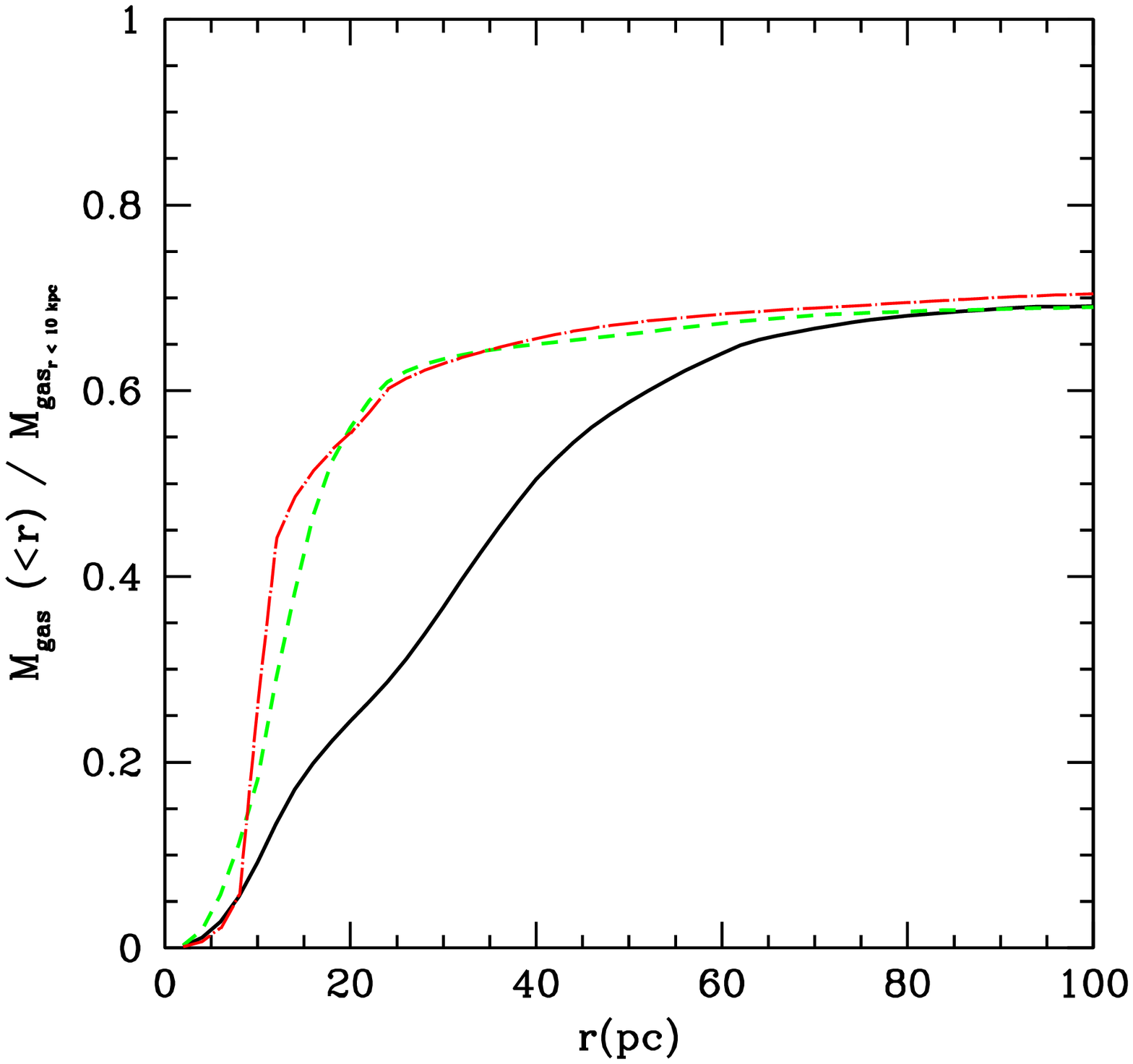}}
\smallskip {\small Figure S2. Cumulative gas mass profile (normalized to the total gas
mass) within the inner $100$ pc. Three simulations with $\gamma=7/5$ for three
different values of the gravitational softening are shown, $40$ pc (black solid line),
$10$ pc (red dot-dashed line) and $2$ pc (green dashed line).}
\end{figure}

The gas fraction, $f_{\rm g}$, is 10\% of the total disk mass. 
The rotation curve of the model is shown in Figure S1.
Different encounter geometries were explored in the large suite of merger simulations
previously performed and published${\it(S15)}$: prograde or retrograde coplanar mergers 
as well as mergers in which one of the disks was inclined with respect to the orbital plane.
The simulation 
presented in this Report is the refined version of a coplanar prograde encounter. This
particular choice is by no means special for our purpose, except that the galaxies
merge slightly faster than in the other cases, thus minimizing the computational
time invested in the expensive refined simulation. We note that the existence
of a coherent nuclear disk after the merger is a general result that does not depend
on the details of the initial orbital configuration, including the initial relative inclination
of the two galaxies${\it(S15)}$. Similarly, gas masses and densities
in the nuclear region differ by less than a factor of 2 for runs having the same initial 
gas mass fraction in the galaxy disks but different initial orbits.

The galaxies approach each other on parabolic orbits with pericentric distances 
that were 20\% of the galaxy's virial radius, typical of cosmological 
mergers${\it(S18)}$ . The initial separation of the halo centers was twice 
their virial radii and their initial relative velocity was determined from the corresponding 
Keplerian orbit of two point masses. Each galaxy consists of $10^5$ 
stellar disk particles, $10^5$ bulge particles, and $10^6$ DM particles. 
The gas component was represented by $10^5$ particles.
We adopted a gravitational softening of $\epsilon = 0.1$ kpc
for both the DM and baryonic particles of the galaxy, and 
for its SMBH $\epsilon=0.03$ kpc.

\subsection{The refined simulations of the nuclear region}

\subsubsection{Particle splitting}

In this Report we use the same technique of particle splitting that we have used before to study the formation 
of a disk galaxy${\it(S19)}$.
Splitting has been already used to follow supermassive
black holes evolving in spherical gaseous backgrounds${\it(S20)}$ and to model the formation of primordial stars and black holes at high redshift${\it(S21)}$.Several schemes for particle splitting have been proposed
, both static and dynamic, and it has been shown that splitting gives robust
results even when simulating highly dynamical systems such as collapsing clouds${\it(S22)}$.
In dynamic splitting the mass resolution is increased during the simulation based on some criterion, 
such as the local Jeans length of the system. This requires extreme care when calculating SPH density or pressure at the boundary between the fine grained and the coarse grained volumes. In static splitting the approach is much more conservative and one simply selects a subvolume to refine. The simulation is then restarted with increased mass resolution just in the region of interest. 
We adopted the latter technique. By selecting a large enough volume for the fine grained region one can 
avoid dealing with spurious effects at the coarse/fine boundary. 
We select the volume of the fine-grained region large enough to guarantee that the dynamical timescale
of the entire coarse-grained region is much longer than the dynamical timescale of the refined region.
In other words, we make sure that gas particles from the coarse region will reach the
fine region on a timescale longer than the actual time span probed in this work. This is important because
the more massive gas particles from the coarse region can exchange energy with the lower mass particles 
of the refined region via two-body encounters, artificially affecting their dynamics and thermodynamics${\it(S20,S23)}$.
Hence our choice to split in a volume of 30 kpc in radius, while the two galaxy cores are separated 
by only $6$ kpc. 
The new particles are randomly distributed according to the SPH smoothing kernel within a volume  of size $\sim h_p^3$, where $h_p$ is the smoothing length of the parent particle. The velocities of the child particles are equal to those of their parent particle (ensuring momentum conservation)
and so is their temperature, while each child particle is assigned a mass equal to $1/N_{\rm split}$ the mass of the parent particle, where 
$N_{\rm split}$ is the number of child particles per parent particle.
The mass resolution in the gas component was originally $2 \times 10^4 M_{\odot}$  
and becomes $\sim 3000 M_{\odot}$ after splitting, for a total of 1.5 million SPH
particles.. The star and dark matter particles are not splitted 
to limit the computational burden.
The softening of the gas particles is reduced to $2$ pc (it was $100$ pc in the low resolution simulations).
For the new mass resolution, the local Jeans length is always resolved by 10 or more SPH smoothing
kernels${\it(S24,S25)}$ in the highest density regions occurring in 
the simulations. The softening of the black holes is also reduced from $30$ pc to $2$ pc.
The softening of dark matter and star particles remains $100$ pc because they are not splitted. Therefore
in the refined simulations stars and dark matter particles essentially provide a smooth background
potential to avoid spurious two-body heating against the much lighter gas particles, while the computation focuses on the gas component which dominates by mass in the nuclear region (see sections 2.1-2.3).
We have verified that, thanks to the fact that gas dominates the mass and dynamics of the
nuclear region, the large softening adopted for the dark matter particles does not affect
significantly the density profile of the inner dark halo that surrounds the nuclear disk.
We constructed an equilibrium gaseous disk embedded
in a dark halo choosing parameters as close as possible to the nuclear disk of the merger remnant
in our simulation. The mass ratio between the equilibrium disk and the dark halo is a factor of 2
lower than that in our simulation in order to facilitate the stability of the disk against 
fragmentation in absence of turbulence (turbulence is a significant stabilizing factor in the 
merger remnant of our standard simulation). We evolve the
system with different particle numbers and choosing a softening of the dark matter particles
as large as the disk radius, as in the refined simulations, or 30 times smaller, i.e. equal to that of 
the gas particles. We compare the different runs after evolving the
gaseous disk for a few orbital times and we find that spurious effects are 
seen in the profiles of the large dark matter softening simulation only at scales 
as small as 3-4 times the softening of the gas particles (Figure S3).
At such distance from the center the density profile flattens out and is about a
factor of 3 lower than in the simulation with small dark matter softening. 
The reason why a flattening of the density profile does not occur 
at a much larger scale of order the dark matter softening
is because gas dominates the inner mass distribution, causing a contraction of the halo  
that overwhelms the tendency to form a constant density core owing 
to the large softening (the effect of halo contraction on the slope of
density profiles is recognized to be important in general and has been
widely studied and demonstrated, even in the context of mergers ${\it(S8)}$).  
Moreover, the total density profile, including the gas component, is even less 
affected (Figure S3). The total density profile is actually most relevant for the  overall strength of dynamical friction. In summary, the effect of the
large dark matter softening is only manifest at scales that approach 
the nominal resolution of the simulation set by the gas softening and therefore it hardly affects the 
sinking of the black hole pair.

\subsubsection{Resolution tests}

In fluid systems for which gravity plays a major role, as it is the case here,
the effective spatial resolution is set by the largest between the gravitational softening and the
SPH smoothing length. These can be forced to be always equal, such as in experiments of molecular cloud
collapse${\it(S26)}$, but this requires introducing adaptive gravitational softening, which can cause
spurious fluctuations in the potential energy of the particles. Here we opt for a fixed gravitational
softening${\it(S8)}$ and we set it in such a way as to have a high force
resolution while being comparable or somewhat larger than the SPH smoothing length (or, more precisely, a spherical volume of radius equal to the softening always contains of order $2*N_{\rm sph}$ or more, where $N_{\rm sph}$ is the
number of neighboring particles used in the SPH calculation (=32 in GASOLINE). 

\begin{figure}
\vskip 11.6cm 
{\includegraphics{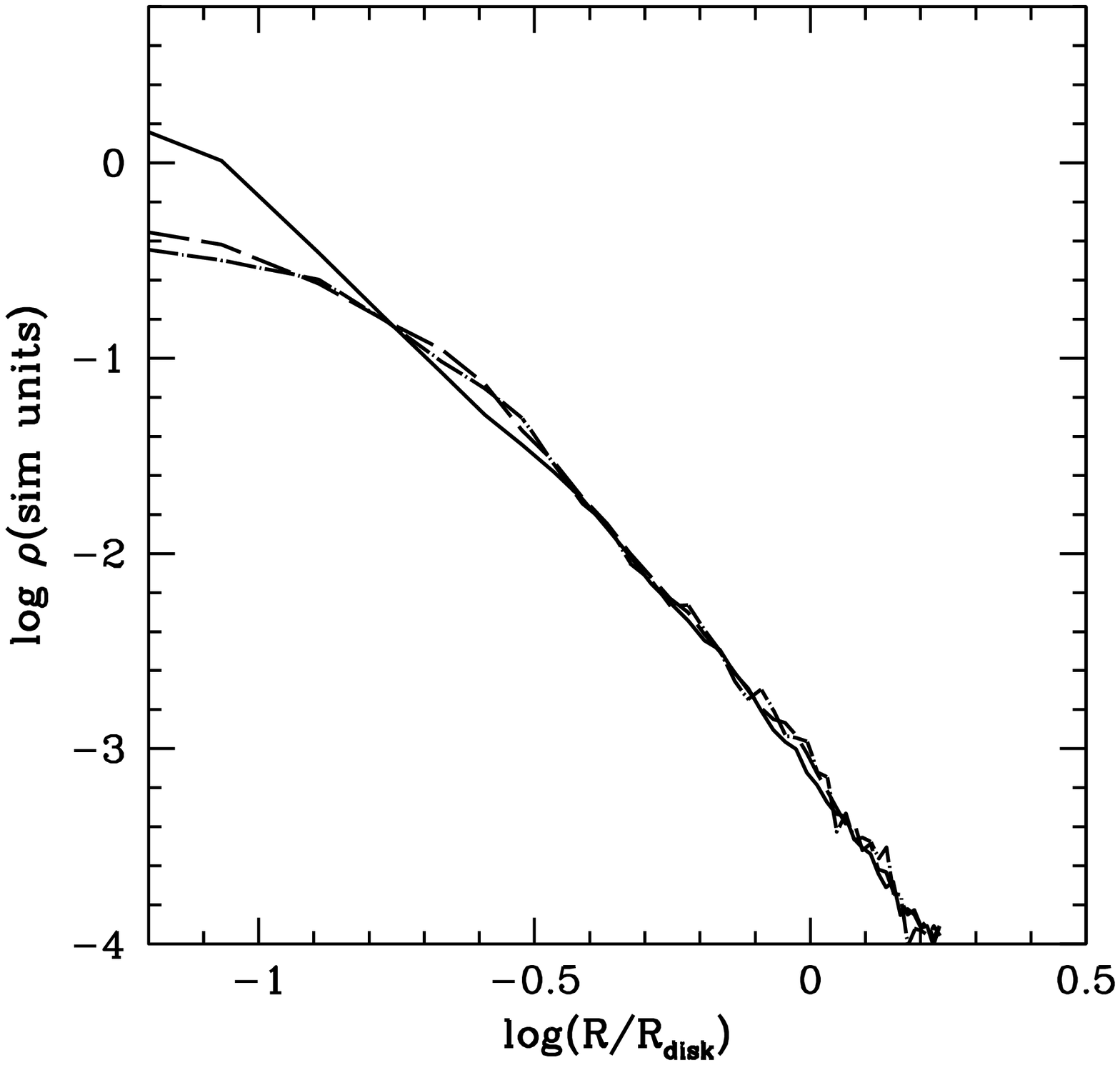}}
{\includegraphics{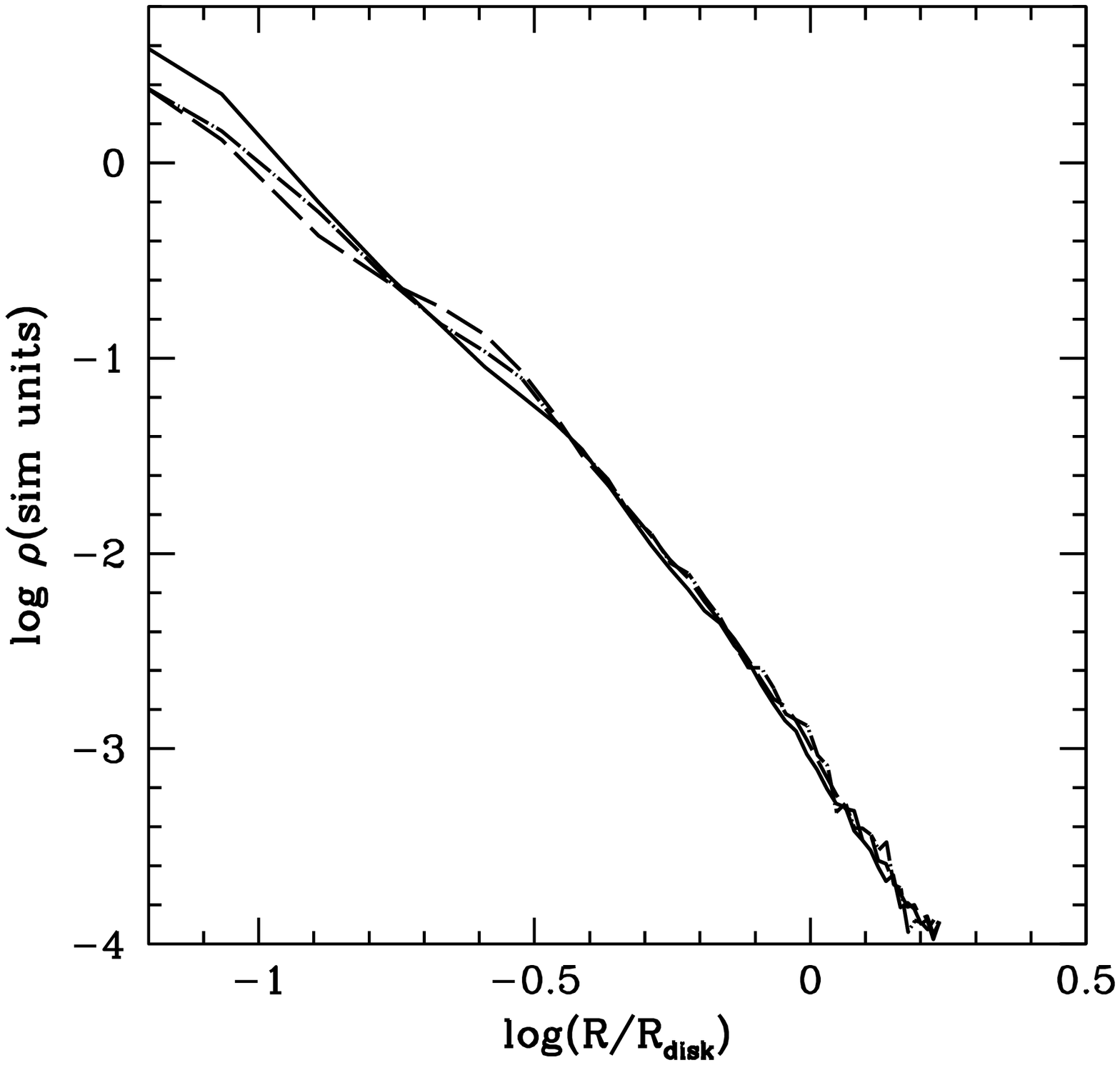}}
\vskip 5 cm
{\small Figure S3. Density profiles of the dark matter (top) and of the sum of the dark matter
and gaseous component (bottom) of an equilibrium nuclear gaseous disk embedded in a dark matter 
halo ($R_{\rm disk}$ is the radius of the nuclear disk and the density is measured in 
simulation units).
The tests were run to study the dependence of the profiles on the softening of
the dark matter component (see 1.3.1 for a description of the test). Different curves correspond to
different simulations. The solid line is used for a simulation with a dark matter softening
$=0.033 R_{\rm disk}$ and $50000$ particles in the halo, the dashed
line refers to a simulation with a dark matter softening equal to $R_{disk}$ and $50000$
halo particles, and the dot-dashed line is used for a simulation with a dark matter
softening equal to $R_{\rm disk}$ and $5000$ halo particles. In all simulations the softening
of the gas is $=0.033 R_{\rm disk}$. The curves are shown from two gas softening lengths outwards.}.
\end{figure}

The latter choice avoids spurious fragmentation${\it(S24,S25)}$.
During the late stage of the merger densities grow considerably, and locally the SPH smoothing lengths
can be appreciably smaller than the softening, with the result that the radial inflow of gas might
be suppressed, as in the analogous case of star forming clouds${\it(S24)}$.
We tested how the results depend on the choice of the gravitational softening of the gas by running the refined part
of the calculation with a softening of $40$ pc, $10$ pc, or $2$ pc (the latter is our
 choice in the reference simulation used in the Report). We adopt an adiabatic equation of state with $\gamma=7/5$
(see next section). We find that the gas mass profile at scales smaller 
than 100 pc, namely in the region of the nuclear disk, approaches convergence at a softening of about $10$ pc
(see Figure S2).
This means that the amount of gas that ends up in the nuclear disk, which ultimately determines the strength of
the drag, is a robust result (see also 2.2). 
Likewise, the scale height and global structural properties of the nuclear disk, 
such as average density, size, sound speed, velocity dispersion and rotational velocity are nearly equal 
when comparing 
the $10$ pc and the $2$ pc runs. This explains why the orbital decay rate is nearly identical until
the two SMBHs are separated by a distance resolved with both the $10$ pc an the $2$ pc simulation${\it(S27)}$. 
On the other end, the non-axisymmetric features in the disk and
the gas inflow at the smallest scales become increasingly better resolved as the gravitational
softening is decreased, implying that an even higher resolution will be necessary to study the fueling of 
the black holes in a robust way.

\subsection{Thermodynamics of the nuclear region: the model}

In the refined simulations the gas is ideal and each gas particle obeys $P=(\gamma - 1) \rho u$. 
The specific internal energy $u$ evolves with time as a result of $PdV$ work and shock heating modeled via  
the standard Monaghan artificial viscosity term (no explicit radiative cooling term is included). We refer 
to section 1.1 for a description of the implementation. 

The entropy of the system increases as a result of shocks. Including irreversible heating from shocks is
important in these simulations since the two galaxy cores undergo a violent collision. 
Shocks are generated even later
as the nuclear, self-gravitating disk becomes non-axisymmetric, developing strong spiral arms.
Therefore the highly dynamical regime modeled here is much different from that considered by previous works 
starting from an equilibrium disk model, which could be evolved using a polytropic equation of state and 
neglecting shock heating${\it(S28-S30)}$.
Radiative cooling is not directly included in the refined simulations. Instead, the magnitude of the
adiabatic index, namely the ratio between specific heats, is changed in order to mimic different 
degrees of dissipation in the gas component, thereby turning the equation of state of the
gas into an ``effective'' equation of state${\it(S26,S31,S32)}$. 

Previous works${\it(S31)}$ have used a two-dimensional radiative transfer code to study the
effective equation of state of interstellar clouds exposed to the intense UV radiation field 
expected in a starburst finding that the gas has an adiabatic index $\gamma$ 
in the range $1.3-1.4 (=7/5)$ for densities in the range $5 \times 10^3-  5 \times 10^4$ atoms/cm$^3$. 
The latter density is comparable to the volume-weighted mean density in our simulated nuclear 
disks. Such values of the adiabatic index are expected for quite a range of starburst intensities, from
$10$ $M_{\odot}$/yr to more than $100$ $M_{\odot}$/yr${\it(S32)}$, hence encompassing the peak star formation rate of 
$\sim 40 M_{\odot}$/yr measured in the original galaxy merger simulations${\it(8)}$. 
Hence under these conditions the nuclear gas is far from isothermal ($\gamma=1$), 
which would correspond to radiative cooling being so efficient to balance heating coming from compression and/or radiative processes, as it happens in the first stage of the simulation. 
The inefficient cooling is mostly due to a high optical depth which causes 
trapping of H$_2$O lines. In addition the warm dust heated by the starburst continuously heats the gas 
via dust-gas collisions, and  the cosmic-rays also heat the gas substantially. 
We adopt $\gamma = 7/5$ in the simulation described in the Report (we have
also run a case for $\gamma=1.3$ and found that the structure of the nuclear disk is substantially
unchanged). In essence, we treat the gas as a one-phase medium whose mean density and internal
energy (the sum of thermal and turbulent energy) will correspond to the mean density and line width 
seen in observed nuclear disks${\it(S33)}$. 

In section 2.2 we provide more details on
the structure of the nuclear disk. For 
densities above $10^5$ atoms/cm$^3$ or below $10^3$ atoms/cm$^3$ cooling is more efficient and 
$\gamma$ should drop even below 1${\it(S31)}$
Therefore, in reality the nuclear disks will have a complex multi-phase structure with temperatures
and densities spanning orders of magnitude, as shown by detailed numerical calculations${\it(S34,S35)}$. 
In particular, the lowest density and highest density gas present in the refined simulation
would be characterized by have an effective sound speed, $v_s = \sqrt{\gamma
k_B T / \mu}$, lower than that in the simulation. A lower gas sound speed will yield
a faster decay of the black hole binary${\it(S28)}$.
This is because the
drag is more efficient in a supersonic rather than in a subsonic regime${\it(S36)}$. Since the
gas in the simulation is already transonic the sound speed need not be much lower for the gas
to become supersonic. A colder gas will also becomes denser, which again goes in the direction
of increasing the sinking rate of the black holes. Hence, if 
anything we err on the side of underestimating the drag by using a constant, high $\gamma$ everywhere. 
A faster decay will only strengthen our scenario. 

We tested that the transition
between the thermodynamical scheme used in the low-res part of the simulation, which adopts the
cooling function described in section 1.1, and the second thermodynamical scheme with the effective
equation of state does not introduce large fluctuations
in the hydrodynamical variables. This was done by rerunning the refined stage of the simulation with the same cooling function 
adopted in the low resolution part of the simulation. This new refined simulation was then compared with the standard $\gamma=7/5$
refined simulation before the merger, just
one crossing time of the inner region (calculated within a volume of $200$ pc around one of the cores) after refinement.
We recall that in the standard simulations we introduce the effective equation of state exactly when we apply the
refinement. The chosen time is shorter than the merging time of the two cores but
long enough to show eventual fluctuations resulting from switching to the
effective equation of state in our standard refined simulation.
We find that the density distributions are nearly identical in the cores, while they are slightly different
in the region of high compression between the two cores. Such differences are of the expected
sign, namely that a larger pressure gradient develops with the effective 
equation of state compared to the cooling simulation as the two disks approach each other (indeed the refined simulation with cooling
behaves almost as an isothermal run, i.e. corresponds to a softer equation of state relative to the $\gamma=7/5$ run).

\begin{figure}
\vskip 11.6cm 
{\includegraphics{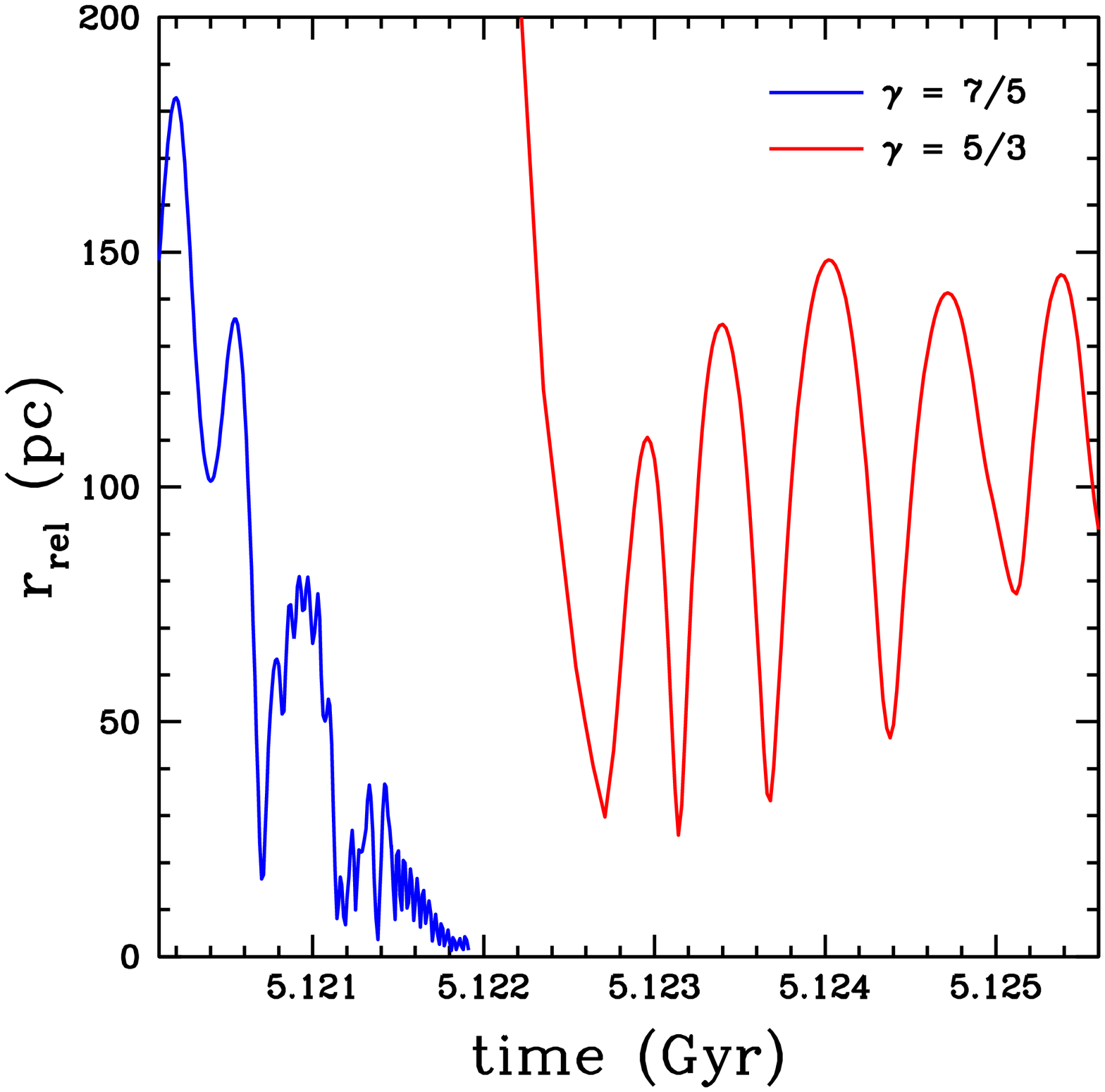}}
{\includegraphics{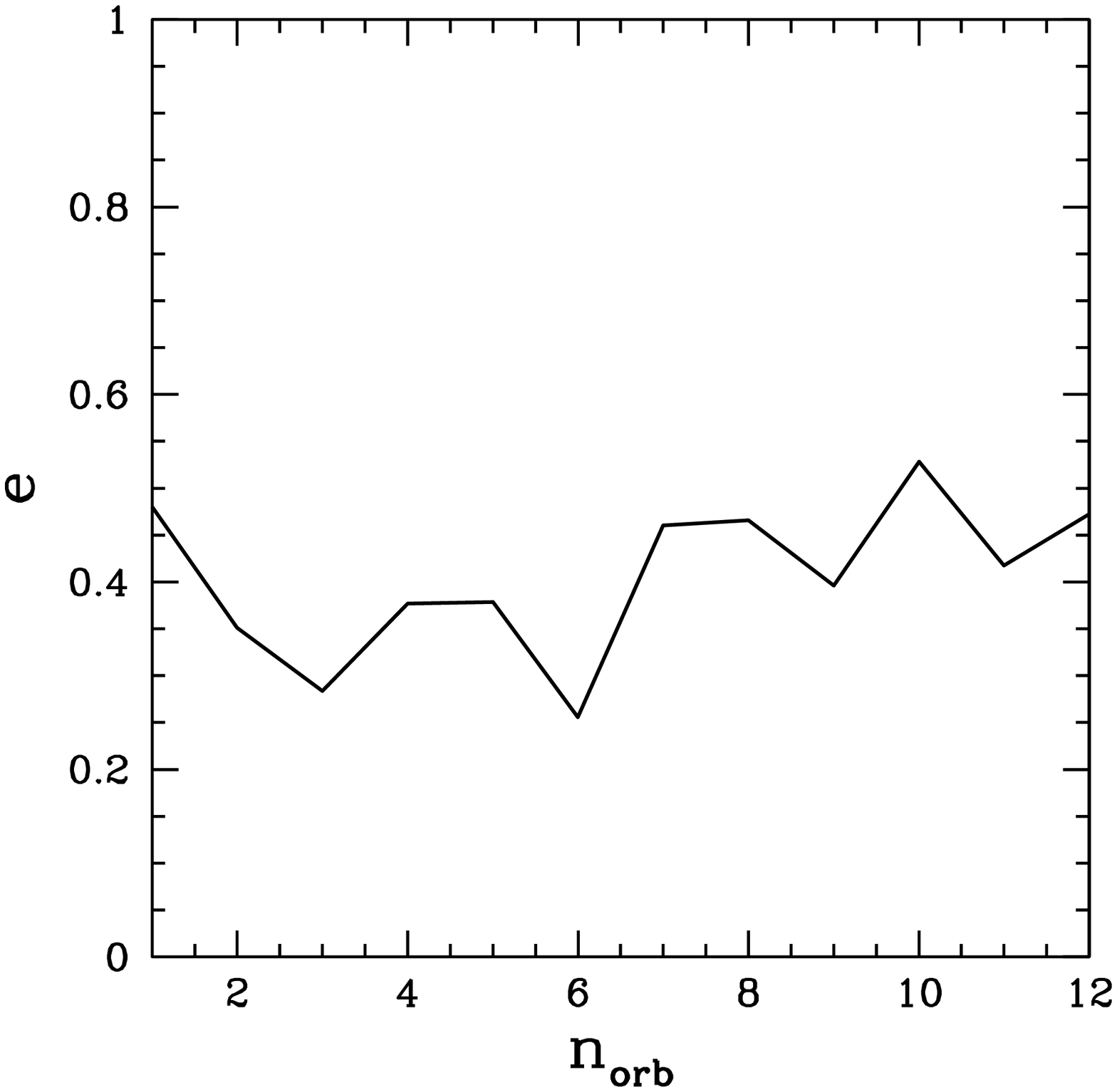}}
\vskip 7cm
{\small Figure S4. 
Top: Orbital evolution of the binary supermassive black holes.
The blue line shows the relative distance as a function of time for the pair in
the $\gamma=7/5$ simulations, as in Figure 2 of the Report, while the red line
shows it for the $\gamma=5/3$ simulation (see section 2.1). Bottom: Evolution of the
eccentricity of the pair of SMBHs as a function of the number of orbits
performed in the nuclear disk in the $\gamma=7/5$ simulation.
The binary forms after about eight orbits.
The time spanned by the orbits corresponds to that in the inset of Figure 2 of the Report.}
\end{figure}

\section{Supporting Online Text}

\subsection{Effects of thermodynamics on the sinking of the SMBHs}

We tested how a smaller degree of dissipation affects the structure and dynamics of the 
nuclear region by increasing $\gamma$ to $5/3$. This would correspond to a purely adiabatic gas,
or equivalently it corresponds to the assumption that radiative cooling is completely negligible. 
The radiative feedback from
an active galactic nucleus (AGN) is a good candidate for a strong heating source that the models
on which we based our prescription for thermodynamics described in the previous section
do not take into account${\it(S31,S32)}$. 
An AGN would not only act as an additional source of radiative
heating but would also increase the turbulence in the gas
by injecting kinetic energy in  the surrounding medium, possibly suppressing gas cooling${\it(S37,S38)}$.
Even before the two galaxy cores merge, when they are still a few kpc away, a mild gas inflow 
collects more than $10^8 M_{\odot}$
within a few hundred parsecs from the black holes. The gas is already arranged in a disk-like
structure, which is highly tidally distorted and presents  inward radial motions.
Therefore, there is room to imagine that significant gas accretion might take place before
the merger is completed, turning one or both the two black holes into an AGN.

\begin{figure}
\vskip 7.0cm 
{\includegraphics{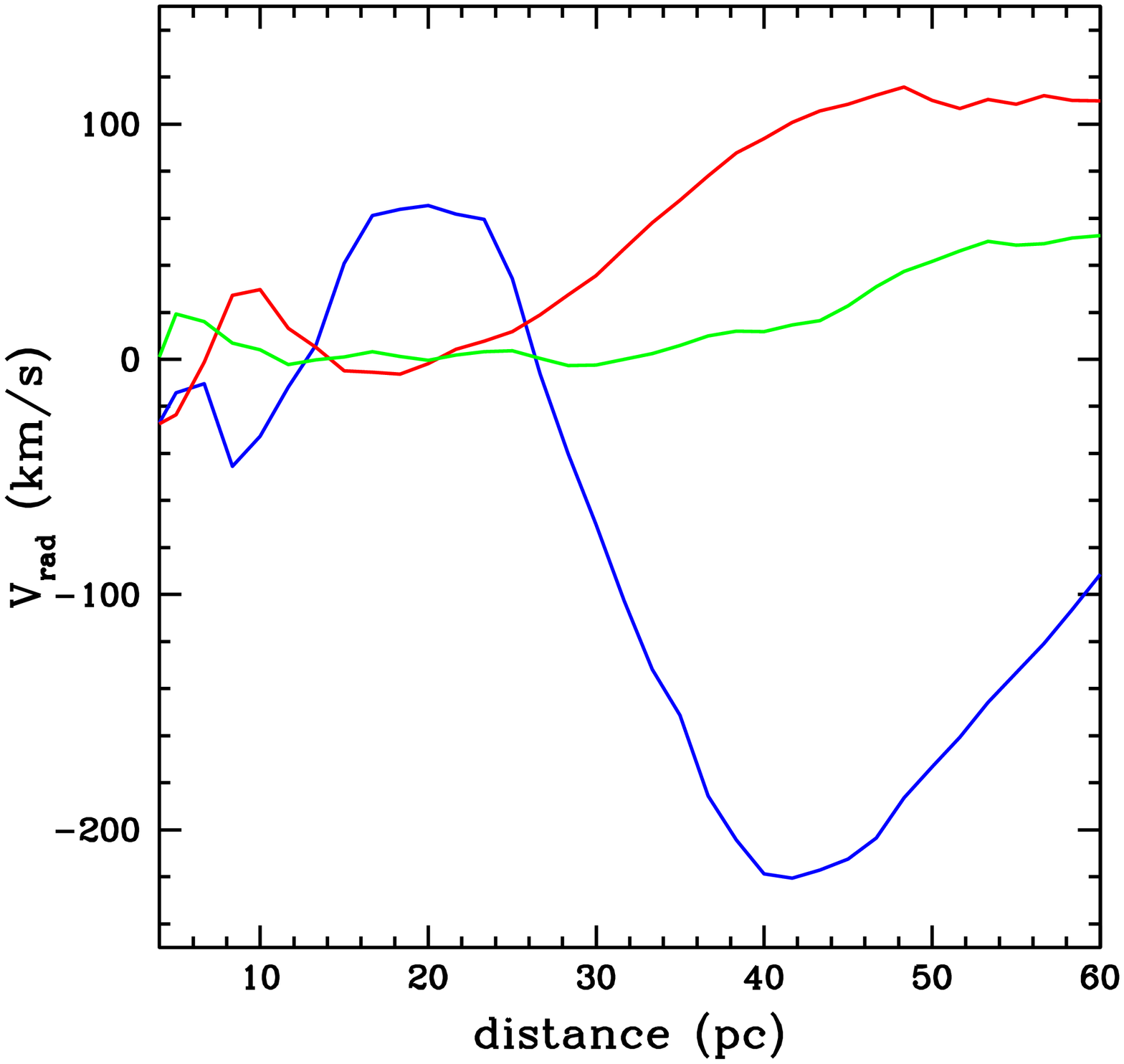}}
\smallskip
{\small Figure S5. Radial velocities within the nuclear disk ($\gamma = 7/5$)
starting at $t=5.1218$ Gyr (blue line), and then after another $10^5$ years 
(red line) and $2 \times 10^5$ years (green line). Remarkable inflow and
outflow regions are the result of streaming motions within 
the bar and spiral arms arising in the disk during the phase of 
non-axisymmetric instability sustained by its self-gravity
At later times the instability saturates due to self-regulation and the radial motions also level
down (green line).}
\end{figure}

We have run another refined simulation with $\gamma=5/3$ to explore this
extreme situation. In this case we find that a turbulent, pressure supported cloud of a few hundred 
parsecs arises from the merger rather than a disk. The mass of gas is lower within 100 pc relative to the 
$\gamma=7/5$ case because of the adiabatic expansion following the final shock at the merging of the cores. The nuclear 
region is still gas dominated, but the stars/gas ratio is $> 0.5$ in the inner $100$ pc.
The black hole pair does not form a binary due to inefficient orbital decay, and maintains a separation 
of $\sim 100-150$ pc (Figure S4) well after they have formed a binary in the $\gamma=7/5$ case. 
The gas is hotter and more turbulent; the sound speed $v_s \sim 100$ km/s and the turbulent velocity 
$v_{\rm turb} \sim 300$ km/s become of the same order of $v_{\rm bh}$, the velocity of the black holes, and the density
around the black holes is $\sim 5$ times lower than in the $\gamma=7/5$ case. The lower density, and to a minor 
extent the fact that the motion of the two black holes occurs in a transonic rather than in a supersonic regime 
greatly reduce the drag due to the gas (see below).

A strong form of AGN feedback which can shut off cooling at galactic as well
as at cluster scales has been sometimes advocated to stop the cooling flow in clusters, create entropy cores 
and, in general, reduce the overcooling problem at large scales${\it(S39,S40)}$.
Given the discovered sensitivity to the value of $\gamma$, it would appear that a scenario in which 
black holes rapidly form a binary and eventually later coalesce owing to the drag provided by the
gas requires that models for AGN feedback be effective on the large scale and simultaneously have negligible thermodynamical effects at small scales. This might indeed be a more general requirement if one has
to preserve the ubiquitous nuclear disk-like structures seen in many merger remnants
(see below section 2.2). Indeed, previous works that included a prescription for AGN feedback
in galaxy merger simulations similar to ours${\it(S37,38)}$ find that feedback affects strongly
the thermodynamics of the gas in the nuclear region only $>10^8$ years after the merger is
completed.

As briefly mentioned in the Report, we have attempted a simple calculation of the accretion rate of the
black holes and of their released energy output assuming Bondi-Hoyle spherical accretion 
of the gas within twice the softening of the black holes, following the scheme
adopted in ${\it(S37,S38)}$.
We have neglected the motion of the black holes relative to the surrounding gas since these
move on orbits corotating with the disk (hence there is nearly no net relative motion between 
the gas and the holes). In addition, we have assumed that the radiative emission is Eddington limited and occurs with an efficiency of $10 \%$ (indeed we find that they can sustain a mean accretion rate of 
$0.05 M_{\odot}$/yr, slightly below Eddington) and
would radiate an amount of energy nearly comparable to the internal energy of the gas (the sum of turbulent, rotational and 
thermal energy) over the time comprised between the formation of the nuclear disk and the formation of the binary. 
The latter result is obtained if all the radiated energy goes into heating the surrounding gas isotropically,
neglecting radiative losses or hydrodynamical instabilities at the interface between different phases, such as Kelvin-Helmoltz 
or Rayleigh-Taylor instabilities, that can convert thermal energy into turbulent energy. Isotropic heating was
also assumed in  ${\it(S37,S38)}$, who indeed matched the $M-\sigma$ relation by depositing only $5\%$ of 
the emitted energy into heating of the gas. If we make the same assumption here then the total radiated
energy over the same timescale is smaller than the internal energy of the gas by more than a factor
of 50. This leads us to conclude that the radiative heating from the SMBHs should not affect significantly 
the thermodynamics of the gas and thus its equation of state.
However, if the black holes begin accreting earlier on, when
the two galaxy cores have not merged yet, the integrated energy output until the binary forms can be much larger 
and could affect the thermodynamics of the newly formed nuclear disk. Indeed once the black holes become active they
would be able to accrete and emit at the Eddington limit for at least a Salpeter time, which 
is $4 \times 10^7$ yr, a timescale almost 100 times longer than the binary formation timescale (there is indeed
enough gas in the nuclear region to sustain an even more prolonged accretion, albeit below Eddington). 
While gas accretion should be more effective, and thus the assumption that emission occurs
at the Eddington limit more sensible, when the nuclear disk has already formed
simply because the gas has larger inward radial motions and thus should feed the
black holes more efficiently (see also section 2.2)
only higher resolution calculations capable of studying in detail the dynamics of the gas near the holes will be able to address the issue of when accretion becomes significant, and if and when it
can produce a significant energetic feedback on the nuclear disk. Furthermore, additional test runs that we have performed (without
black holes) show that if the equation of state becomes stiffer ($\gamma=5/3$ instead of $\gamma=7/5$) only {\it after}
the merger is completed, the nuclear disk structure is barely 
affected (this would correspond to the case in which the SMBHs become active only after the formation of the nuclear disk). 
Therefore the timing of accretion during the merger is a crucial aspect that will have to be explored by future work. Section 2.2 will briefly discuss how the structure of the inner gaseous distribution in runs
with different values of $\gamma$ compares with the observations of the nuclear regions of 
merger remnants.

In the $\gamma=5/3$ case the black holes could still decay and form a binary as a result of the
interaction with the stellar background. Since the resolution of the stellar background in
the simulations is likely inadequate to assess directly the effect of dynamical friction,
we calculated the dynamical friction timescale in a collisionless background analytically${\it(S41)}$ 
using

\begin{eqnarray}
{\tau_{DF}=1.2 {
V_{cir}r_{cir}^2
\over[GM_{bh}/{\rm{e}}]\ln(M_{sd}/M_{bh})}\,\varepsilon^{0.4}} 
\end{eqnarray}

where $V_{\rm cir}$ and $r_{\rm cir}$ are, respectively, the initial
orbital velocity and the radius of the circular orbit with the
same energy of the orbit of the black holes in the simulation, and $\varepsilon$ is the circularity of the
orbit ($\varepsilon = J(E)/J_{cir}(E)$, where $J(E)$ is the angular momentum of the orbit
as a function of its orbital energy $E$  and $J_{circ}(E)$ is the angular momentum of a circular
orbit having the same orbital energy $E$, so that $\varepsilon =0$ corresponds to a radial orbit and
$\varepsilon =1$ corresponds to a circular orbit), $M_{sd}$ is the sum of the
dark matter and stellar mass within $r_{cir}$. We calculate the decay time
starting from when the two black holes are $100$ pc apart, namely at the
periphery of the nuclear disk, just after the merger. Drawing the numbers
from the simulation we set $r_{\rm cir} = 100$ pc, $V_{\rm circ}$= 200 km/s,  $\varepsilon =0.5$,
$M_{\rm bh} = 2.4 \times 10^6 M_{\odot}$ and $M_{\rm sd} = 5 \times 10^8 M_{\odot}$. 
We find that timescales for dynamical friction timescale in a collisionless background are 
$5 \times 10^7$ yr and $3 \times 10^7$ in the $\gamma=5/3$ and $\gamma=7/5$ case, respectively
(the shorter timescale in the $\gamma=7/5$ case is due to the fact that the stars and halo adiabatically 
contract more in response to the higher gas mass concentration, hence $M_{\rm sd}$ is higher).
In comparison, the binary formation timescale in the $\gamma=7/5$ simulation is only $5 \times
10^5$ years.  

Equation 1  was derived for an isothermal sphere. The stellar and dark matter 
distribution are indeed only mildly triaxial within a few hundred parsecs from the center ($c/a > 0.7$, where $c$ is the semi-minor axis and $a$ the semi-major axis of the mass distribution) and the total profile is extremely 
close to that of an isothermal sphere${\it(S8)}$. The fact that the merger remnant
is not far from spherical confirms the predictions of larger-scale simulations regarding the effect of
gas cooling on the structure of the global potential${\it(S42,S43)}$.
Note that equation actually yields a lower limit to the dynamical friction timescale since close to parsec scales, as the binary becomes hard, evacuation of the stellar background due to three-body encounters
will begin, and the efficiency of the sinking process will be greatly reduced. Whether sinking will
continue and eventually lead to coalescence of the two holes is uncertain in this case given the fact
that the gas does not play an important role. Centrophilic orbits in triaxial systems would help in refilling the
loss cone, and could in principle bring the black holes down to the distance where gravitational waves
would take over${\it(S44)}$. However, as we just mentioned, the structure of the stellar core
is only mildly triaxial. Further investigation with simulations having higher resolution in the collisionless
component is needed. The $\gamma=5/3$ run was stopped $5 \times 10^6$ years after the merger of the gaseous cores 
is completed. Once again, the fact that there is no evidence that the holes are sinking until the end is
likely due to insufficient mass and force resolution in the collisionless background that does
not allow to resolve dynamical friction properly${\it(S45)}$ (as we will explain in the remainder of this
section the gas contributes little to the drag in this case).

We also compared our results with the expected dynamical friction timescale due to the gaseous 
background. In the run with $\gamma=7/5$ the gas is distributed in a disk rather than in
an isothermal sphere. Since the disk thickness is $> 10$ times the black hole gravitational 
softening and owing to the fact that the density profile of the disk can be roughly 
approximated with a power law
with an index close to 2 (except at the center where it becomes steeper)  we are allowed to use 
eq. (1) to obtain a rough estimate. 
As previously shown${\it(S28)}$, analytical
predictions${(S36)}$ can overestimate the drag in the supersonic regime by a factor $\sim 1.5$. 
In the $\gamma=7/5$ case the regime is mildly supersonic and the analytical formula should yield the
correct prediction. In this case the drag is $\sim 2.3$ time stronger than in the
corresponding collisionless case${\it(S28)}$. This is fairly consistent with our results. Indeed, formula (1) with a reduction of a factor of $2.3$ would give $2.3 \sim 10^6$ yr if we set $M_{gas}=M_{sd}$, the 
gas mass being about 20 times more than the mass in stars. This timescale has to be compared with 
that measured in the simulation, $5 \times 10^5$ yr (we recall that the gas profile is steeper
than $r^{-2}$ near the center, therefore it is not surprising that the decay is actually faster). 

Despite the apparent agreement with the analytically estimated drag we note that the orbital dynamics
of the two black holes might be affected by more than just the gravitational wake. Indeed the
disks show strong , highly dynamical non-axisymmetric features such as spiral arms (see next section); 
in the analogous, well-studied case of planet migration orbital decay is well described by torques 
exerted onto the target body by the spiral modes and its efficiency depends on the location of resonances
between the orbital motion and the spiral pattern that extract or deposit angular momentum${\it(S46)}$.
The latter description of the orbital decay might be more appropriate here rather 
than just considering the effect of the gravitational wake (see also the discussion on the orbital
eccentricity below)

The drag drops rapidly by an order of magnitude approaching the subsonic
regime${\it(S20)}$; this coupled with the fact that $M_{gas}$ is a factor of 5 lower in the $\gamma=5/3$ would give
a drag 50 times smaller in the latter case, explaining why the orbital decay provided by the gas is so 
inefficient in such conditions.

In summary, in the $\gamma=7/5$ run the sinking timescale due to the
gas is much shorter than that due to the stellar background because of the combination 
of (1) the fact that gas densities are much higher than stellar densities in the center and (2) the 
fact that in the supersonic regime the drag in a gaseous background is stronger than that in a stellar 
background with the same density.  Adding star formation is unlikely to change this conclusion. In fact the 
low-resolution galaxy mergers simulations yield a starburst timescale of $5 \times 10^7$ yr. During this
time, which is much longer than the binary formation timescale, half of the gas in the nuclear 
disk is turned into stars. Instead,  in the $\gamma=5/3$ stars and gas would
contribute to the drag in a comparable way as the black holes begin to sink
but since the sinking timescale is much longer and
comparable with the star formation timescale the overall orbital evolution will be dictated
by the stars rather than by the gas.

There are, however, some caveats in our argument regarding the role of star formation in the $\gamma=7/5$
case. First, the starburst timescale is based on the low resolution merger simulations. Had we included
star formation in the refined simulations we would have probably found shorter timescales locally
since these simulations are capable of resolving much higher densities and the star formation rate 
depends on the local gas density. Second, one might wonder how the inclusion of feedback from star
formation, which was neglected in the low resolution merger simulations,  would affect gas properties 
and, eventually, the orbital decay of the black holes. As for the first issue, we can obtain a rough
estimate of how short the star formation timescale can be in the following way. We note that in the
nuclear disk most of the gas is at densities above $100$ atoms/cm$^3$. At these densities molecular
hydrogen formation is efficient ${\it(S47)}$.
Let us then make the rather extreme assumption that all the gas in the disk is molecular and
readily available for star formation. Then, let us simply assume that
molecular gas will be turned into stars on the local orbital timescale. Star formation in molecular clouds
is rather inefficient, and typically 30\% of the dense, molecular gas only is converted into stars, possibly
because internal turbulence in the clouds prevents them from collapsing altogether  ${\it(S48)}$. Therefore let us
write the star formation rate in the nuclear disk as a whole as $dM_*/dt = 0.3 \times M_{gas}/T_{orb}$,
where $T_{orb} = 10^6$ years, the orbital time at the disk half mass radius, $25$ pc, and 
$M_{gas} = 3 \times 10^9 M_{\odot}$.
The resulting star formation rate is $900 M_{\odot}$/yr , about 25 times higher than that estimated 
in the low-res simulations. Nonetheless, even with such high star formation rate less than 1/5
of the gas in the disk, $4.5 \times 10^8 M_{\odot}$, would be converted into stars during the time 
required for the black holes to sink and bind in the nuclear disk ($5 \times 10^5$ years). 
Regarding the issue of feedback, radiative feedback from stars is implicitly included in our choice 
of the equation of state in the $\gamma=7/5$ case (see above), but feedback from supernovae 
explosions is not taken into account.
However, supernovae feedback would contribute to both heating the gas and increasing
its turbulence, which should go in the direction of decreasing the star formation rate and therefore
strengthening our previous argument concerning the role of star formation. 
Moreover, while it will have remarkable effects on the
multi-phase structure of the gas $({\it S34-S35})$, it should not have a major impact on the
energetics of the disk in the $\gamma=7/5$ case. In fact, assuming
a star formation rate of $900 M_{\odot} / yr$ and a Miller-Scalo initial
stellar mass function we obtain that 
supernovae should damp $\sim 4 \times 10^{51}$ erg/yr ($7 \times 10^{48}$ erg per solar mass of stars
formed) into the surrounding gas, corresponding to   $\sim 2 \times 10^{57}$ erg damped
during the binary formation timescale, $5 \times 10^5$ yr.
This is about $50\%$ smaller than the internal energy of the gas in the nuclear disk 
(the sum of turbulent, rotational and thermal energy). 
However, since the decay of the black holes will be sensitive to changes in the local gas density along their orbit, only future calculations that incorporate directly the effects of star formation and 
supernovae explosions will probably find quantitative differences relative to our simple thermodynamical 
model.

We note the black holes sink on an eccentric orbit in the nuclear disk (see above). We tracked the
evolution of the eccentricity after the two cores merge and the holes are embedded in a single disk and
found that despite fluctuations the eccentricity $e$ after about $10^6$ yr is roughly identical to its initial value, $e \sim 0.5$ (Figure S4).
Only in the initial phase of the decay we do signs a tendency of circularization. This is different from
what found in ${\it(S30)}$, who measured fast circularization of for binaries of SMBHs evolving in equilibrium
nuclear disks. A caveat in the comparison is that we can only follow a few orbits once the binary is 
formed in the refined simulations. Nevertheless, a tendency towards circularization is clear already
during the first few orbits in ${\it (S30)}$, contrary to what we find here. The reason for this difference 
is not clear
but it is probably related to differences in the structure of the nuclear disk. The disk in the refined
simulation has a much stronger spiral pattern than that in ${\it(S30)}$ due to its much higher self-gravity,
having a mass $30$ times higher.
Analytic calculations of tidal torques in the context of planet migration  ${\it(S49)}$ have shown
that a massive body moving in an eccentric 
disk is characterized by an orbit whose eccentricity is comparable to the degree of non-axisymmetry  of the disk. 
Such calculations considered a simple m=1 spiral mode superimposed on the disk potential
as the source of the ``eccentricity'' of the disk (the mode is treated as a forcing term in the
equations of motion) but 
the results should be quite general in a qualitative sense. The consequence is that a black hole in a strongly non-axisymmetric disk should move on a more eccentric orbit relative to the case of a mildly axisymmetric disk, and that this tendency 
for the natural orbit (in the sense that it is the natural solution of the equations of motion)
to be eccentric should counteract the tendency of dynamical friction to circularize the orbit
as the black holes decay. The net outcome will depend on how the structure of the nuclear disk
evolves with time. This is an important issue that warrants further investigation 
because the coalescence time of the binary in the gravitational radiation dominated phase
depends on its orbital eccentricity.

Finally, the last stage of the refined simulations provides the initial conditions for future
models that will eventually follow the hardening of the binary below parsec scales. These future calculations
will be able to show whether the last parsec problem can be overcome in the nuclei of merger 
remnants with gas, as suggested by simulations of binary SMBHs embedded in equilibrium nuclear disks
${\it(S20, S28-S30)}$.
The binary should continue to sink as a result of gas drag . Its sinking rate, however,
 will strongly 
depend on the dynamics and thermodynamics of the disk at scales below the resolution of 
the refined simulations. The properties of the gas below one parsec
will determine whether a gap will be opened by the binary, slowing down significantly albeit not
stopping its orbital decay, or whether the holes will sink at a faster rate as a result of torques
by the asymmetries in the mass distribution surrounding the two holes ${\it(S28-S30)}$.
Whether or not an appreciable fraction of the gas will be converted into stars by the
time the separation of the binary has fallen below a parsec will also have
an impact since, for instance, the conditions for gap formation will depend on the 
local gas density and the overall sinking rate might have a non-negligible contribution
from the stars.
 
\subsection{Structure and kinematics of the nuclear disks}

The nuclear disk produced in the $\gamma=7/5$ case is highly turbulent. The source of turbulence
are the prominent shocks generated as the cores merge and the persistent non-axisymmetric structure 
sustained by the self-gravity of the disks after the merger is completed${\it(S34,S35)}$.
The perturbation due to the binary black holes is a negligible effect since their mass is about $10^3$ times 
smaller than the mass of the disk (we tested this by restarting a simulation after 
removing the black holes once the merger is completed). The degree of turbulence,
of order $50-100$ km/s as measured by the radial velocity dispersion, is comparable to that of
observed circumnuclear disks${\it(S33,S47)}$. 
The disk is composed by a very dense, compact region of size about 25 pc which contains
half of its mass (the mean density inside this region is $> 10^5$ atoms/cm$^3$). The
outer region instead, from 25 to 75-80 pc, has a density 10-100 times lower, and is surrounded
by even lower density rotating rings extending out to a few hundred parsecs.  The disk scale
height also increases from inside out, ranging from 20 pc to nearly 40 pc.
The volume-weighted density within 100 pc is in the range $10^3-10^4$ atoms/cm$^3$, comparable to 
that of observed nuclear disk ${(\it S33)}$.
This suggests that the degree of dissipation implied by our equation 
of state is a reasonable assumption despite the simplicity of the thermodynamical scheme adopted.

The rotating, flattened cloud produced in the $\gamma=5/3$ is instead more turbulent and less
dense than observed circumnuclear disks in merger remnants. The mean velocity dispersion measured within
$100$ pc is about $300$ km/s, higher than the mean rotational velocity within the same radius,
which is $\sim 250$ km/s. This suggests that the  $\gamma=5/3$ simulation does not describe the
typical nuclear structure resulting from a dissipative merger,
The strong spiral pattern in the disk produces remarkable radial velocities. Since spiral modes transfer angular momentum inwards and mass outwards${\it(S52,S53)}$
, strong inward radial velocities are present. The amplitude
of radial motions evolves with the amplitude of the spiral pattern, in the sense that radial motions
decline as the spiral arms weaken over time.
Just after the merger, when non-axisymmetry
is strongest, radial motions reach amplitudes of $\sim 100$ km/s (Figure S5). 
This phase lasts only a couple of orbital
times, while later the disk becomes smoother as spiral shocks increase the internal energy which in turn
weakens the spiral pattern.
Inward radial velocities of order $30-50$ km/s are seen for the remaining few orbital times
during which we are able to follow the system (Figure S5). Such velocities are comparable to those recently
seen in high resolution observations of the nuclear disk of nearby Seyfert galaxies${\it(S54)}$.
As the gas reaches down to a few parsecs from the center 
its radial  velocity diminishes because one approaches the limits of the gravitational force resolution in the
simulation ($\sim 2$ pc). Therefore the fact that there is almost no net radial velocity within a 
few parsecs from the center (Figure S5)
is an artifact of the limited resolution. In addition, in this innermost region the gas is so dense ($\rho
> 10^5$ cm$^{-3}$) that our equation of state breaks down, since a lower $\gamma$, close to the isothermal value,
would be more appropriate (see section 1.4). A higher dissipation rate likely means the gas inflow rate in the innermost region
of the disk will be higher than the one seen in the current simulations${\it(S55)}$.  If we assume that speeds of 
$30-50$ km/s can be sustained down
to scales of a few parsecs, $> 10^8 M_{\odot}$ of gas could reach parsec scales in about $10^5$ yr. The latter timescale is
much smaller than the duration of the starburst, and therefore such gas inflow should develop in a similar way even
when star formation is taken into account. The inflow is also marginally faster than the decay timescale of the binary SMBH measured in the simulations, which is $\sim 5 \times 10^5$ years. Presumably some of this gas 
could be intercepted by the two SMBHs as they are spiraling down (the relative velocities between the gas and the 
black holes are small since the SMBHs are always corotating with the nuclear disk), but the higher dissipation
rate expected near the central, denser regions 
might increase the magnitude of the drag due to the gas in reality and bring the sinking time of the
binary very close to the gas inflow timescale. In the latter case the two SMBHs could undergo massive accretion,
and probably become active while they are sinking. Only future simulations that include a realistic treatment
of the gas thermodynamics at all scales will be able to study gas accretion and orbital decay simultaneously. 

Our single-phase simulations seem to describe reasonably well the average properties of observed
nuclear disk because they capture the thermodynamics of the low density, pressurized
medium that has the largest volume filling factor ${\it(S33)}$.
Yet they do not model directly the molecular gas, which will have temperatures of order
$100$ K, much lower than the effective temperature produced by our equation of state,
and comprises most of the gas mass in observed nuclear disks  ${\it(S33)}$.
The dense molecular phase would be better described
as an ideal gas with $\gamma=1$ or lower (see 1.4). While a model of the multi-phase interstellar
medium in the disk should be the ultimate goal for future simulations, one could also design an
intermediate scheme in which the equation of state (in particular $\gamma$) changes as a function of
density, as done in simulations of molecular cloud collapse ${\it(S26)}$.
If a large fraction of the gas has a much softer equation of state compared to the $\gamma=7/5$
runs a much denser and thinner disk will likely form. We performed a simulation that
adopts an isothermal equation of state ($\gamma = 1$) in the refined part of the calculation 
and indeed the resulting disk has a mass comparable to that in the previous calculations
but a scale height of only $\sim 5$ pc,
which would probably shrink even further with a higher spatial resolution. The sound speed
is only $10$ km/s compared to $60$ km/s in the $\gamma=7/5$ case, and turbulence is also
much weaker. The disk is violently 
gravitationally unstable, with dense rings and arms, which would likely fragment with a smaller
softening. The simulation was stopped after the merger because the very high densities at the center
prevented an efficient integration. However, the cores merged faster than in the $\gamma=7/5$ case
and we expect the black holes would also sink faster based on the results of previous work ${\it(S28)}$
that found how the orbital decay is faster when the gas has a higher density and/or
has a lower sound speed, even in the case in which the disk is significantly clumpy.
In reality the nuclear region will have the global properties, especially
the global energetics, of our $\gamma = 7/5$ simulation, while locally the densest gas 
will have properties closer to that of the isothermal test run. Such a complex multi-phase
structure is predicted by simulations of self-gravitating turbulent disks ${\it(S34-S35)}$.

\bigskip

{\large{\bf References and Notes}}

S1.  Wadsley, J., Stadel, J., \& Quinn, T., {\it New Astr}., {\bf 9}, 137 (2004)

S2. Barnes, J., \& Hut, P., {\it Nature}, {\bf 324}, 446 (1986)

S3. Gingold, R.A. \& Monaghan, J.J., {\it Mon. Not. R. Astron. Soc.}, {\bf 181}, 375 (1977)

S4. Monaghan, J.J., {\it Annual. Rev. Astron, Astrophys}, {\bf 30}, 543 (1992)

S5. Barnes, J., {\it Mon. Not. R. Astron. Soc.},  {\bf 333}, 481 (2002)

S6. Springel, V. and Hernquist, L., {\it Mon. Not. R. Astron. Soc.}, {\bf 333}, 649 (2002)

S7. Balsara, D.S., {\it J.Comput. Phys.},121, 357 (1995)

S8. Kazantzidis, S., Mayer, L., Colpi, M., Madau, P., Debattista, V., Quinn, T., Wadsley, J. 

\hspace{+12pt}\& Moore,  B., {\it Astrophys. J.}, \textbf{623}, L67 (2005)

S9. Katz, N. {\it Astrophys. J.}, {\textbf 391}, 502 (1992)

S10. Governato, F., Mayer, L., Wadsley, J., Gardner, J. P., Willman, B., Hayashi, E., Quinn, 

\hspace{+12pt}T., Stadel, J.\& Lake, G, {\it Astrophys. J.}, {\bf 607}, 688 (2004)

S11. Hernquist, L., {\it Astrophys. J. Supp.}, {\bf 86}, 389 (1993)

S12. Springel, V. \& White, S.D.M., {\it Mon. Not. R. Astron. Soc.}, {\bf 307}, 162 (1999)

S13. Mo, H. J., Mao, S., White, S. D. M., {\it Mon. Not. R. Astron. Soc.}, {\bf 295}, 319 (1998)

S14. Navarro, J.~F., Frenk, C.~S. \& White, S.~D.~M., {\it Astrophys. J.}, \textbf{462}, 563 (1996)

S15. Kazantzidis, S., Mayer, L., Mastropietro, C., Diemand, J., Stadel, J., \& Moore, B., 

\hspace{+12pt}{\it Astrophys. J.}, {\bf 608}, 663 (2004)

S16. {Klypin} A.,  {Zhao} H. \&   {Somerville} R.~S., {\it Astrophys. J.}, {\bf 573}, 597 (2002)

S17. Blumenthal, R.G., Faber, S.M., Flores, R., \& Primack, J.R., {\it Astrophys. J.}, {\textbf 301}, 27 

\hspace{+12pt}(1986)

S18. Khochfar, S. \& Burkert, A., {\it Astron. Astrophys.}, \textbf{445}, 403 (2006)

S19. Kaufmann, T., Mayer, L., Wadsley, J., Stadel, J. \& Moore, B., {\it Mon. Not. R. Astron}. 

\hspace{+12pt}{\it Soc.}, \textbf{370}, 1612 (2006)

S20. Escala, A., Larson, R. B., Coppi, P. S., \& Mardones, D., {\it Astrophys. J.}, \textbf{607}, 765 (2004)

S21. Bromm, V. \& Loeb, A., {\it Astrophys. J.}, \textbf{596}, 34 (2002)

S22. Kitsionas, S. \& Whitworth, S., {\it Mon. Not. R. Astron. Soc.}, \textbf{330}, 129 (2002)

S23. Steinmetz, M. \& White, S.D.M., {\it Mon. Not. R. Astron. Soc.},  \textbf{288}, 545 (1997)

S24. Bate, M. \& Burkert, A., {\it Mon. Not. R. Astron. Soc.}, \textbf{288}, 1060 (1997)

S25. Nelson, A.F., {\it Mon. Not. R. Astron. Soc.},  \textbf{373}, 1039 (2006)

S26. Bate, M.R., Bonnell. I.A., \& Bromm, V., {\it Astrophys. J.}, \textbf{336}, 705 (2002)

S27. Mayer, L., Kazantzidis, S., Madau, P., Colpi, M., Quinn, T., \& Wadsley, J., 2006b, 

\hspace{+12pt}Proc. Conf. Relativistic Astrophysics and Cosmology - Einstein's Legacy (astro-ph/0602029).

S28. Escala A., Larson, R. B., Coppi, P. S. \& Mardones, D., {\it Astrophys. J.}, \textbf{630}, 152 (2005)

S29.  Dotti, M., Colpi, M. \& Haardt, F., {\it Mon. Not. R. Astron. Soc.}, \textbf{367}, 103 (2006)

S30. Dotti, M., Colpi, M., \& Haardt, F., \& Mayer, L., {\it Mon. Not. R. Astron. Soc.}, in press 

\hspace{+12pt}(astro-ph/0612505)

S31. Spaans, M. \& Silk, J.,{\it  Astrophys. J}, \textbf{538}, 115 (2000)

S32. Klessen, R.S., Spaans, M., Jappsen, A., {\it Mon. Not. R. Astron. Soc.}, \textbf{374}, L29 (2007)

S33. Downes, D. \& Solomon, P. M., {\it Astrophys. J.}, \textbf{507}, 615 (1998)

S34. Wada, K., {\it Astrophys. J.}, \textbf{559}, L41 (2001)

S35. Wada, K. \& Norman, C., {\it Astrophys. J},  \textbf{566}, L21 (2002)

S36. Ostriker, E., {\it Astrophys. J.}, \textbf{513}, 252 (1999)

S37.  Springel, V., Di Matteo, T., \& Hernquist, L.{\it Mon. Not. R. Astron. Soc.}, \textbf{361}, 776 (2005)

S38. Di Matteo, T., Springel, V. \& Hernquist, L., {\it Nature} \textbf{7026}, 604 (2005)

S39. Croton, D. J. {\it et al.}, {\it Mon. Not. R. Astron. Soc.}, \textbf{365} 11 (2006)

S40. Bower, R.G., {\it et al.}, {\it  Mon. Not. R. Astron. Soc.}, \textbf{370}, 645 (2006)

S41. Colpi, M., Mayer, L. \& Governato, F., {\it Astrophys. J.}. {\textbf 525}, 720 (1999)

S42. Dubinski, J.,{\it Astrophys. J.}, {\bf 431}, 671 (1994)

S43. Kazantzidis, S., Kravtsov, A.V., Zentner, A.R., Allgod, B., Nagai, D. \& Moore, 

\hspace{+12pt}{\it Astrophys. J.}, {\textbf 611}, L73-L76 (2004)

S44. Beczik, P., Merritt, D., Spurzem, R. \& Bischof, H.P., {\it Astrophys. J.}, {\bf 641},  L21 (2006)

S45. Weinberg, M. \& Katz, N., {\it Mon. Not. R. Astron. Soc.}, {\bf 375}, 425 (2007)

S46. Papaloizou, J.C.B., \& Larwood, J.D., {\it Mon. Not. R. Astron. Soc.}, {\bf(315}, 823 (2000)

S47. Schaye, J. {\it Astrophys. J.}, {\bf 809}, 667 (2004)

S48. Li, Y., Mac Low, M.M., \& Klessen, R.S., {\it Astrophys. J.}, {\bf 626}, 823 (2005)

S49. Papaloizou, J.C.B., {\it Astron. Astrophys.}, {\bf 388}, 615, (2002)

S50. Greve, T.R., Papadopoulos, P.P., Gao, Y., \& Radford, S.J.E., submitted to Astrophys. 

\hspace{+12pt}J. (2006) (astro-ph/0610378)

S51. Laughlin, G., Korchagin, V. \& Adams, {\it Astrophys. J.}, {\bf 477}, 410 (1997)

S52. Mayer, L., Quinn, T., Wadsley, J. \& Stadel, J.,{\it Astrophys. J.}, {\bf 609}, 1045 (2004)

S53. Fathi, K. et al. {\it Astrophys. J.}, \textbf{641}, L25 (2006) 

S54. Escala, A., {\it Astrophys. J.}, {\bf 648},  L13 (2006)

\end{document}